\documentclass[9pt,shortpaper,twoside,web]{ieeecolor}
\usepackage{generic}
\usepackage{cite}
\usepackage{amsmath,amssymb,amsfonts}
\usepackage{algorithm,algpseudocode}
\usepackage{graphicx}
\usepackage{textcomp}
\usepackage{xspace}
\usepackage{url}
\usepackage[utf8]{inputenc}
\usepackage[english]{babel}
\usepackage{lipsum}
\usepackage{dblfloatfix}
\let\labelindent\relax
\usepackage[margin=2cm]{geometry}
\usepackage [english]{babel}
\usepackage{amsmath, amssymb}
\usepackage{enumitem, array}
\usepackage{subfigure}
\usepackage{tablefootnote}
\usepackage{xcolor}
\usepackage{colortbl}
\usepackage[utf8]{inputenc}
\usepackage[english]{babel}
\usepackage{multicol}
\usepackage{multirow}
\usepackage{tikz}
\usepackage{booktabs}
\newcommand*\emptycirc[1][1ex]{\tikz\draw (0,0) circle (#1);} 

\newcommand*\fullcirc[1][1ex]{\tikz\fill (0,0) circle (#1);}
\usepackage[colorinlistoftodos]{todonotes}
\usepackage{soul}
\newcommand{\remove}[1]{}

\def\BibTeX{{\rm B\kern-.05em{\sc i\kern-.025em b}\kern-.08em
    T\kern-.1667em\lower.7ex\hbox{E}\kern-.125emX}}
\begin{document}
\title{Efficient Query Execution on Encrypted Genotype-Phenotype Database }
\title{PrivGenDB: Efficient and privacy-preserving query executions over encrypted SNP-Phenotype database}
\author{Sara Jafarbeiki, Amin Sakzad, Shabnam Kasra Kermanshahi, Raj Gaire, Ron Steinfeld, Shangqi Lai, Gad Abraham 
\thanks{S. Jafarbeiki, A. Sakzad, R. Steinfeld and S. Lai are with the Faculty of Information Technology, Monash University, Melbourne, Australia. S. Jafarbeiki and R. Gaire are with CSIRO Data 61, Australia. S. Kasra Kermanshahi is with the School of Computing Technologies, RMIT University, Melbourne, Australia. G. Abraham is with the Systems Genomics Laboratory, Baker Institute, Melbourne, Australia (e-mails: sara.jafarbeiki@monash.edu, amin.sakzad@monash.edu; ron.steinfeld@monash.edu; shangqi.lai@monash.edu, raj.gaire@data61.csiro.au, shabnam.kasra.kermanshahi@rmit.edu.au, gad.abraham@baker.edu.au).}
}

\maketitle

\newcommand{\DBGePh}{\mathsf{GDB}}
\newcommand{\Setup}{\mathsf{Setup}}
\newcommand{\GePh}{\mathsf{\mathrm{G} \phi}}
\newcommand{\DOID}{\mathsf{ID}_O}
\newcommand{\DO}{O}
\newcommand{\Trustee}{\mathcal{T}}
\newcommand{\Vetter}{\mathcal{V}}
\newcommand{\Server}{\mathcal{D}}
\newcommand{\SNP}{\mathsf{G}_s}
\newcommand{\Phenotype}{\mathsf{G}_\rho}
\newcommand{\Metadata}{\mathsf{G}_\Delta}
\newcommand{\DB}{\mathsf{DB}}
\newcommand{\g}{g}
\newcommand{\G}{\mathsf{G}}
\newcommand{\User}{\mathcal{U}}
\newcommand{\GSet}{\mathbb{S}_{\G}}
\newcommand{\GInv}{\mathsf{Inv}_{\G}}
\newcommand{\GePhOXT}{\mathsf{\mathbb{G}e \mathbb{P}h OXT}}
\newcommand{\gtoken}{\mathcal{G}\mathsf{token}}
\newcommand{\stag}{\tau_{\rho}}
\newcommand{\gTok}{\mathsf{gToK}}
\newcommand{\gtag}{\tau_{\G}}
\newcommand{\inv}{\mathsf{inv}}
\newcommand{\rrho}{\varrho}
\newcommand{\K}{\mathsf{K}}
\definecolor{mygray1}{gray}{0.6}
\definecolor{mygray2}{gray}{0.8}

\definecolor{orange}{cmyk}{0.1,0.5,0.9,0.3}

\begin{abstract}
Searchable symmetric encryption (SSE) has been used to protect the confidentiality of genomic data while providing substring search and range queries on a sequence of genomic data, but it has not been studied for protecting single nucleotide polymorphism (SNP)-phenotype data.
In this article, we propose a novel model, PrivGenDB, for securely storing and efficiently conducting different queries on genomic data outsourced to an honest-but-curious cloud server. 
To instantiate PrivGenDB, we use SSE to ensure confidentiality while conducting different types of queries on encrypted genomic data, phenotype and other information of individuals to help analysts/clinicians in their analysis/care. 
To the best of our knowledge, PrivGenDB construction is the first SSE-based approach ensuring the confidentiality of shared SNP-phenotype data through encryption while making the computation/query process efficient and scalable for biomedical research and care. Furthermore, it supports a variety of query types on genomic data, including count queries, Boolean queries, and $k'$-out-of-$k$ match queries. Finally, the PrivGenDB model handles the dataset containing both genotype and phenotype, and it also supports storing and managing other metadata like gender and ethnicity privately. \remove{the PrivGenDB model not only can handle the dataset containing both genotype and phenotype, but it also supports storing and managing other metadata like gender and ethnicity privately.}Computer evaluations on a dataset with $5,000$ records and $1,000$ SNPs demonstrate that a count/Boolean query and a $k'$-out-of-$k$ match query over $40$ SNPs take approximately $4.3$s and $86.4\mu$s, respectively, outperforming the existing schemes.
\end{abstract}

\begin{IEEEkeywords}
Searchable symmetric encryption, genomic data privacy, secure outsourcing.
\end{IEEEkeywords}

\section{Introduction}
\label{sec:introduction}

Advances in genotyping technology have enabled the genotyping of large human cohorts and biobanks, (e.g., UK Biobank \cite{UKbiobank1,UKbiobank2}, NIH All of Us \cite{NIH}, the Million Veteran Program \cite{MVP}, and others).
Such large datasets are critical for advancing our understanding of human health and disease and their genetic components, for improving treatment practices, and for better delivery of healthcare. Specific examples include diagnosing a genetic predisposition to disease and precision medicine (e.g., predict the outcome of a specific drug or treatment) \cite{personalized1,personalized2}.
In addition to these large projects, hospitals and research institutions often collect similar data (on a smaller scale) and store them in local databases with limited resources, while the volume of data can be very large.

Single nucleotide polymorphisms (SNPs) are the most common type of genetic variation, and many genetic analyses involve identifying associations between SNPs and traits or diseases.
There are some useful queries/analyses that can help genome-based researchers with their studies, e.g., SNP-disease association study. 
Examples of such queries include count queries, adding negation terms to the query predicates, and sequence similarity search. Other queries such as Boolean query and $k'$-out-of-$k$ match can also help clinicians in medical care.

One of the key challenges with such large datasets is managing and maintaining them securely while enabling appropriate access
to authorised individuals. One solution is the use of cloud computing.
However, cloud-based servers can be vulnerable to cyber breaches, and the information residing on the servers may be exploited by an adversary capable of violating security \cite{cloudsecurity1,cloudsecurity2,cloudsecurity3}.
Therefore, privacy preservation remains an essential factor in the development of electronic frameworks for storing/managing genomic data \cite{privacy1,privacy2}.

Balancing the privacy and security of sensitive information with the need to share and analyse information for research purposes is challenging.
While analysing genomic data is valuable to researchers, 
the increased availability of such data has major consequences for individual privacy; 
notably, because the genome has some essential features, such as (i) connection with individual characteristics and certain disorders, (ii) identification capability, and (iii) disclosure of family relationships \cite{naveed2015privacy}.
Measures need to be taken to ensure this data stays in the right hands.
With more sensitive information (genomic information), the need to preserve privacy rises considerably.


Traditional ways of privacy protection for medical data, such as de-identification, are often not sufficient. 
The Health Insurance Portability and Accountability Act of 1996 (HIPAA) \cite{HIPAA} provides guidelines 
about how to handle sensitive information through de-identification. 
To maintain privacy during data processing and storage, as well as during subsequent data analysis tasks, organisations also consider government policies and regulations \cite{BigDataforHealth}.
Nowadays, most de-identification approaches struggle to defend 
against re-identification attacks, and lead to utility loss \cite{berger2019emerging, gymrek2013identifying}. Therefore, cryptographic solutions are more attractive for privacy-preservation.\par
\definecolor{magentaa}{cmyk}{0.2,0.4,0,0}
\begin{table*}
\caption{Comparison of existing schemes for query execution on genomic data}
\label{table:first comparison}
\centering
\resizebox{15cm}{!}{%
\begin{tabular}{ccccccccccc}
\toprule
\multirow{2}{*}{Scheme}&\multirow{2}{*}{Method}&\multirow{2}{*}{Primitive}&\multirow{2}{*}{Solution}&
\multicolumn{2}{c}{Functionality}&\multirow{2}{*}{}&\multicolumn{3}{c}{Privacy}&\multirow{2}{*}{Efficiency$^\S$}\\
\cline{5-6}\cline{8-10} &&&&Dataset&Queries&&Data&Query&Output&\\
\midrule
\cite{kantarcioglu2008cryptographic}  & PE & Asymmetric & Software & SNP & C &&\fullcirc[0.7ex] &\emptycirc[0.7ex] &\fullcirc[0.7ex]& $1260$ s \\
\cite{ghasemi2016private} & PE & Asymmetric&Software & SNP,PH &C,top $k$ &&\hspace{1.3mm}\fullcirc[0.7ex]* &\fullcirc[0.7ex] &\fullcirc[0.7ex]& $29$ s \\
\cite{nassar2017securing} & PE & Asymmetric&Software & SNP &C,N,$k'$ & &\fullcirc[0.7ex] &\emptycirc[0.7ex] &\fullcirc[0.7ex]& $\sim10$ s $^1$ \\

\cite{hasan2018secure} & PE,GC & Asymmetric&Software & SNP,PH$^+$ & C & &\fullcirc[0.7ex] &\fullcirc[0.7ex] &\fullcirc[0.7ex]& $130.8$ s \\
\cite{chenghong2017scotch} & PE,SGX & Asymmetric & Hybrid & SNP,PH & C & &\fullcirc[0.7ex] &\fullcirc[0.7ex] &\fullcirc[0.7ex] & $1.475$ s $^2$\\
\cite{canim2011secure} & SCP,AES & Symmetric & Hybrid & SNP,PH & C & &\fullcirc[0.7ex] &\emptycirc[0.7ex] &\fullcirc[0.7ex] & $20$ s\\
{\cite{new2018}} & {PE,GC} & Asymmetric&Software & SNP,PH & C & &\fullcirc[0.7ex] &\fullcirc[0.7ex] &\fullcirc[0.7ex]& $31$ s $^3$\\
\cite{new2020} & PE,GC & Asymmetric&Software & SNP,PH$^+$,M & C & &\fullcirc[0.7ex] &\fullcirc[0.7ex] &\fullcirc[0.7ex]& $163.92$ s\\
\remove{\rowcolor{magentaa}\cite{new2021} & PE & Asymmetric&Software & SNP,PH$^+$ & C & &\fullcirc[0.7ex] &\fullcirc[0.7ex] &\fullcirc[0.7ex]& $\sim70$ s\\}

PrivGenDB & SSE & Symmetric&Software & SNP,PH$^+$,M & C,N,$k'$,B & &\fullcirc[0.7ex] &\fullcirc[0.7ex] &\hspace{1.3mm}\fullcirc[0.7ex]* & $1.3$ s\\
\bottomrule
\multicolumn{11}{p{17cm}}{Notations:
PE: Paillier Encryption; AES: Advanced Encryption Standard; GC: Garbled Circuit; SCP: Secure Cryptographic Coprocessor; SGX: Software Guard Extensions; SSE: Searchable Symmetric Encryption; Hybrid: Software \& Hardware, PH: Phenotype; M: Metadata; C: Count; N: Negation; $k'$: $k'$-out-of-$k$ match; B: Boolean; $^+$: Phenotype includes the exact disease/trait, not just positive/negative signs to show whether that record has one particular phenotype or not; *: Security guarantee is met with some leakage. $^\S$: Efficiency is compared in terms of query execution time, over $5,000$ records and $1000$ SNPs, with query consisting of $10$ SNPs; $\sim$:Approximate number extracted from figures in related papers; $^1$:This is an approximate time for $5,000$ records and $300$ SNPs; $^2$:This time is for just $173$ records and $1000$ SNPs with query consisting of 25 SNPs; $^3$:Time for $100,000$ SNPs and less than $173$ records.}\\
\end{tabular}
}
\end{table*}
Several existing cryptographic solutions have been utilised in different schemes \cite{hasan2018secure,kantarcioglu2008cryptographic,ghasemi2016private,nassar2017securing,chenghong2017scotch,canim2011secure,new2018,new2020} for encrypting genomic data and running queries on this encrypted data.
Moreover, following privacy patterns can be defined when outsourcing genomic data and conducting different queries against it: Data Privacy (DP), Query Privacy (QP), and Output Privacy (OP) \cite{hasan2018secure} (see definitions in Section \ref{label:privacyrequirements}).
Table \ref{table:first comparison} summarises the method, cryptographic primitive, solution type, types of dataset used, supported queries, and privacy considerations taken into account by these schemes. 



As illustrated in the table, previous schemes do not execute a set of valuable queries (count, Boolean for retrieving data, $k'$-out-of-$k$ match, negation) altogether
on encrypted genomic data efficiently and they are not scalable to benefit cutting-edge biomedical applications. Moreover, they provide limited support for storing and managing dataset containing additional information (like gender and ethnicity) apart from the genomic data and phenotypes privately.
Some of those solutions are only software-based ones, while others are hybrid, meaning they exploit both software and hardware. None of the previous schemes used a software-based method which is symmetric that works relatively fast and offers low computation complexity. We hypothesise that recent developments in searchable symmetric encryption can offer more functionality efficiently.\par

Searchable encryption is a cryptographic technique which allows searching over encrypted data. 
It has two key research dimensions: searchable symmetric encryption (SSE) and public key encryption with keyword search (PEKS) \cite{SEsurvey}. 
SSE has obtained much more attention from the researchers and evolved well with more expressive searching and higher efficiency \cite{SSE20181,SSE20182,SSE2016,SSE2020,SSE2019new,SSE2021new}. 
SSE has also been studied to protect genomic data for different purposes/functionalities, such as enabling substring search \cite{substringsearch}, or range query \cite{rangequery} on encrypted sequences of genomic data. Nevertheless, it has not been studied for preserving SNP-phenotype data privacy. 
One of the most significant celebrated SSE methods  is Oblivious Cross Tag (OXT), introduced in \cite{cash2013highly}. OXT enables searches over encrypted data in the form of Boolean queries efficiently. Our work proposes a novel model, PrivGenDB, which utilises SSE scheme motivated by OXT, to provide privacy, functionality, and scalability.


\subsection{Our Contributions}
We present a new model, which we call PrivGenDB, for providing privacy of outsourced genomic data stored on the cloud supporting data confidentiality and query privacy. Our model also provides output privacy for all the sensitive information retrieved from the server. (see Table \ref{table:first comparison}-Privacy). PrivGenDB is the first model that provides the privacy patterns without relying on hardware (e.g., using SGX) and is based on symmetric cryptographic primitives only.
Inspired by OXT scheme of \cite{cash2013highly}, we construct the first secure and efficient SSE scheme to instantiate PrivGenDB. We also prove that PrivGenDB achieves the same security guarantees as those in \cite{cash2013highly}. To facilitate such a construction, we present a novel encoding method for genomic data, and construct an inverted index based on that encoding. Hence, our model can handle datasets containing genotype and phenotype (the exact disease) and other information like gender and ethnicity (see Table \ref{table:first comparison}-Dataset). 
In terms of functionality, our PrivGenDB enables us to handle rich queries on genomic data (see Table \ref{table:first comparison}-Queries), such as count query, Boolean query, $k'$-out-of-$k$ match query and negation predicates (see Section \ref{queries} for detailed information about these queries).\remove{(i) count query (aggregate query): counting the number of matches between the query keywords and the stored data, (ii) Boolean query: retrieving IDs of patients that have a specific genomic data or disease or etc, (iii) $k'$-out-of-$k$ match query: checking if the number of matched keywords reaches the number of specified query keywords, (iv) negation predicates: determining whether or not a certain SNP is associated with a given genotype.} We evaluate the computational costs (of the initialisation and search phases), storage size, and communication costs of our PrivGenDB scheme and compare these against the state-of-the-art schemes. We have conducted several experiments and the results show the effectiveness of our SSE-based PrivGenDB in comparison with previous works, which were mainly based on the Paillier Encryption algorithm. For example, for a count query on a dataset of $5,000$ records and $1,000$ SNPs, it takes approximately $130.8$ s in \cite{hasan2018secure}, while PrivGenDB improves the query execution time to $1.3$ s only.

 \newtheorem{rmk}{Remark}
 \newtheorem{pf}{Proof}
 \newtheorem{simulator}{Simulator}
 \newtheorem{pot}{Proof of Theorem \ref{thm2}}

\section{Preliminaries}
\remove{In this section, we provide the required preliminaries, including a biology background regarding genomic/phenotypic databases and their used queries, as well as some cryptographic definitions.}
\subsection{Biology Background}
This section provides background on genomic data, phenotype, and some common queries against this type of dataset.
\subsubsection{Genomic data and Phenotype}
The human genome consists of DNA (deoxyribonucleic acid) with two long complementary polymer chains of four units called nucleotides,
A (Adenine), C (Cytosine), G (Guanine), and T (Thymine). 
Single nucleotide polymorphism (SNP) is a position in the genome where more than one base (A, T, C, G) is observed in a population. Most SNPs are biallelic and only two possible variants (\emph{alleles}) are observed. The set of specific alleles carried by an individual is called its \emph{genotype}.
Together, SNPs account for a large proportion of the genetic variation underlying many human traits such as height and predisposition to disease (also called \emph{phenotype}). 


\remove{One of the most common genetic analyses is the genome-wide association study (GWAS), which aims to identify which SNPs are associated with the trait in question, and thus help understand the genetic mechanisms underlying these traits (also called \emph{phenotype}).
}

Table \ref{table:Data Representation with SNPs} illustrates one type of data representation in a database of SNP genotypes for a number of individuals \cite{canim2011secure,hasan2018secure}, together with the individuals' phenotype.
\remove{Each row reflects the genotype and phenotype for one individual. Each SNP's genotypes are represented in a single column \cite{canim2011secure,hasan2018secure}.} \remove{The last column in Table \ref{table:Data Representation with SNPs} represents the phenotype (in this case, different cancer types have been stored as different phenotypes for each record).}
\begin{table}[htb]
\centering
\caption{Data Representation with SNPs} \label{table:data with snp}
\scriptsize

\begin{tabular}{ccccccc}
\hline
\toprule
\textbf{Record} & {\textbf{SNP$_{1}$}} & {\textbf{SNP$_{2}$}} & {\textbf{SNP$_{3}$}} & {\textbf{SNP$_{4}$}}  & {...} & {\textbf{Phenotype}} \\ \midrule
1                          & AG                        & CC                        & TT                        & AG                                                & …                        & Cancer A                       \\ \hline
2                          & AA                        & CC                        & CT                        & AG                                               & …                        & Cancer B                       \\ \hline
3                          & AG                        & CT                        & CC                        & AA                                               & …                        & Cancer A                       \\ \hline
4                          & AG                        & CC                        & TT                        & AG                                               & …                        & Cancer C                       \\ \hline
5                          & AA                        & CC                        & TT                        & AG                                               & …                        & Cancer B                       \\ \hline
6                          & AA                        & CC                        & TT                        & GG                                               & …                        & Cancer A                       \\ \hline
7                          & AG                        & CT                        & CT                        & AG                                                & …                        & Cancer B                       \\ \remove{\hline
8                          & AA                        & CC                        & TT                        & AG                                               & …                        & Cancer B                       \\ \hline
9                          & GG                        & CT                        & CT                        & AG                                               & …                        & Cancer A                       \\ \hline
10                         & AG                        & CT                        & CT                        & AG                                               & …                        & Cancer A  \\} \bottomrule
\end{tabular}
\label{table:Data Representation with SNPs}
\end{table}

\subsubsection{Queries on Genomic Data}\label{queries}
While biobanks contain individuals' genomic data, different types of queries need to be performed on the data to help researchers analyse and correlate diseases and genome variations such as SNPs. 
Here are some types of queries on genomic data:
\paragraph{Count Query}\label{section:secure count query}
\remove{One of the most common queries for researchers to do statistical analysis is count query. For example, doctors/clinicians are interested in mining the potential variations for certain diseases. This is obtained by a statistical test, which is performed by computing a series of count queries \cite{chen2017princess}.}
A count query determines the number of records that match the query predicates in the database and helps in computing several statistical measurements for genetic association studies.\remove{It is used in disease susceptibility to test for rare variants within a patient’s DNA sequence.} 
An example of a count query on the dataset represented in Table \ref{table:Data Representation with SNPs} is as follows: \par
\emph{Select count from sequences where 
{{$\mathsf{diagnoses=Cancer~B}$ AND $\mathsf{SNP_{2}=CC}$ AND $\mathsf{SNP_{4}=AG}$}}}



The total number of SNPs specified in the query predicate is called the query size. In the above example, the query size is 2, and the answer to the mentioned query is 2 (i.e., records 2 and 5\remove{satisfy the query predicates}).\footnote{For count query: researchers can count the number of records with a specific genotype(s) for a given SNP(s) while having a particular disease.\remove{Hence, they can use the information for statistical analysis.} Furthermore, some other categorisation can also be added to the query,\remove{like asking for the number of people who have mentioned characteristics,} e.g., if they are male/female or even in a certain ethnicity or nationality group.}

\par

\paragraph{Boolean Query}
\remove{Apart from counting the number of matched ones, sometimes we need to extract the data owners' information to help them with some medication or inform them about some drug allergy or vaccine.}
A query from clinicians can be about identifying patients who have specific genotypes, so that they can contact them to provide personalised medication or inform them about their predisposition for drug allergy or vaccine. This can be done by extracting the IDs whose genotypes match the query symbols, then using those IDs to extract other information needed by the clinician. \footnote{Of course, access control has to be considered since this type of information can be released to the clinicians, not the analysts (Note that we do not discuss access control mechanisms in this paper; however, they have been extensively studied in the literature).}
An example of a Boolean query on the dataset represented in Table \ref{table:Data Representation with SNPs} is as follows:\par
\emph{Retrieve IDs from sequences where 
{{$\mathsf{diagnoses=Cancer~B}$ AND $\mathsf{SNP_{2}=CC}$ OR $\mathsf{SNP_{4}=AG}$}}}

The answer to the above query is a list of records' IDs that match the predicates and is: \{2,5,7\}. 

\paragraph{$k'$-out-of-$k$ match Query}
Sometimes the answer to this query can be yes/no when the query symbols are being checked for a particular record. In this case, we can check if the SNPs for a particular record has the query predicates' characteristics or not by some threshold defined.
In our scenario, presented in algorithms in Section \ref{construction}, the IDs related to records who have this threshold matching can also be retrieved to check their other information like diseases or medications.
Examples of a threshold query on the dataset represented in Table \ref{table:Data Representation with SNPs} are as follows:\par
\begin{itemize}
\item  \emph{Does record number $7$ have at least $k'=2$ matches out of $k=3$: 
{$\mathsf{ID=7}$ AND $\mathsf{SNP_{2}=CC}$ AND $\mathsf{SNP_{3}=CT}$ AND $\mathsf{SNP_{4}=AG}$}}.
If this ID satisfies the query predicates, the answer will be the ID, which means yes. Otherwise, it will return $0$, which means no. Answer: yes.\par
\item  \emph{Retrieve IDs from sequences where at least $k'=2$ matches out of $k=3$: {$\mathsf{diagnoses=Cancer~B}$ AND $\mathsf{SNP_{2}=CC}$ AND $\mathsf{SNP_{3}=CT}$ AND $\mathsf{SNP_{4}=AG}$}}.
The answer to this query is a list of records' IDs that match the first predicate and have at least 2 of the rest predicates.
Answer: \{2,5,7\}.\par
\end{itemize}
\paragraph{Negation predicates}
In all of the above queries, negation terms can be added as predicates. For instance,

\emph{select count from sequences where
{{$\mathsf{diagnoses=Cancer~B}$ AND $\mathsf{SNP_{2}\neq CC}$ AND $\mathsf{SNP_{4}=AG}$}}}. Answer: 1 (Record 7 satisfies the query predicates).


\subsection{Cryptographic Background}
This section reviews the cryptographic primitives we opted for in our proposed model.\par

\subsubsection{Searchable Symmetric Encryption (SSE)}\label{OXT}
One significant work in the area of Searchable Encryption is the scheme proposed by Cash et al. \cite{cash2013highly} called Oblivious Cross Tag (OXT), which is important due to its functionality as it supports conjunctive search and general Boolean queries. OXT provides a solution for searches over multiple keywords in the form of general Boolean queries. 
A special data structure called tuple-set or TSet is used for single keyword searches. 
An additional dataset called XSet is used for multiple keyword searches to verify if the identified documents for the first keyword also satisfy the remaining query. The OXT scheme achieves sublinear search time even though the search query is of a Boolean type and not just a single keyword. This time is proportional to the number of documents with the least common keyword $\mathsf{sterm}$ \cite{cash2013highly}.
%
The syntax of SSE is as follows:
\begin{description}
\item $\mathsf{EDBSetup(\lambda, DB)}$: Given security parameter $\lambda$ and a database $\mathsf{DB}$, this algorithm outputs the encrypted database $\mathsf{EDB}$.
\item $\mathsf{Search(q, EDB)}$: Inputs the search query $\mathsf{q}$ and the encrypted database $\mathsf{EDB}$, then outputs the search result (This protocol first uses the TokenGeneration algorithm to generate tokens and then perform the search on encrypted database).
\end{description}
\remove{
\subsubsection{Syntax of Oblivious Cross Tags (OXT)}
The proposed protocol by Cash et al. \cite{cash2013highly}
called Oblivious Cross Tags (OXT) is one of the most important SSE works, which has been used widely in literature and different applications. This scheme supports Boolean queries on encrypted data in sublinear time. Let $\lambda$ be the security parameter. A database $\mathsf{DB=(ind_i, w_i)}$ is a list of (identifier, keyword) pairs. A query $\mathsf{q(\bar{w})}$ is specified by a tuple of keywords and a Boolean formula on $\mathsf{\bar{w}}$. The syntax of OXT SSE is as follows:}
\remove{
\begin{description}
\item $\mathsf{OXT.Setup(\lambda, DB)}$: Given security parameter $\lambda$ and a database $\mathsf{DB}$, it outputs the encrypted database $\mathsf{EDB}$ and the keys $\mathsf{K}$.
\item $\mathsf{OXT.TokenGeneration(w, K)}$:  If the client wants to make a query $\mathsf{q}$ over $\mathsf{EDB}$, search tokens are required. This algorithm generates search tokens based on the given query.
\item $\mathsf{OXT.Search(Tok_q, EDB)}$: inputs the search token $\mathsf{Tok_q = (stag, xtoken[1], xtoken[2], \ldots)}$ for a query $\mathsf{q}$ and $\mathsf{EDB}$, then outputs the search result $\mathsf{ERes}$.
\item $\mathsf{OXT.Retrieve(ERes, K)}$: This algorithm inputs the encrypted search result $\mathsf{ERes}$ and the utilised key, then outputs the documents identifier $\mathsf{ind}$.
\end{description}
}
We review the TSet instantiaition below since we use it in our construction, and we refer the interested readers to the full paper \cite{cash2013highly} for the complete OXT syntax and more details.

\textbf{TSet Instantiation}. Cash et al. instantiate a $\mathsf{TSet}$ in \cite{cash2013highly} as a hash table with $B$ buckets of size $S$ each.  A database $\mathsf{DB=(ind_i, w_i)}$ is a list of (identifier, keyword) pairs. Based on the total number $N = \Sigma_{ \mathsf{w} \in \mathsf{W}} |\mathsf{T}[\mathsf{w}]|$ of tuples in $\mathsf{T}$, the setup procedure lets the parameters $B$ and $S$ in such a way that after storing $N$ elements in this hash table, the likelihood of an overflow of any bucket is a sufficiently small constant; and the total size B.S of the hash table is $O(N)$. It consists of the following algorithms:
\begin{description}
\item 
$\mathsf{TSet.Setup(T)}$: This algorithm takes as input an array $\mathsf{T}$ of lists of equal-length bit strings indexed by the elements of $\mathsf{W}$, in which for each $\mathsf{w}\in \mathsf{W}$, $\mathsf{T}[\mathsf{w}]$ is a list  $\mathsf{t} =(s_1,\ldots, s_{\mathsf{T}_\mathsf{w}})$ of strings. ($\mathsf{T}[\mathsf{w}]$ contains one tuple per each DB document which matches $\mathsf{w}$, i.e., $\mathsf{T_w}$ = $\mathsf{|DB(w)|}$. It outputs a pair $\mathsf{TSet, K_T}$.
\item $
\mathsf{TSet.GetTag(K_T, w)}$:  This algorithm takes $\mathsf{K_T}$ and $\mathsf{w}$ as inputs and outputs $\mathsf{stag}$.
\item $
\mathsf{TSet.Retrieve(TSet, stag)}$: This algorithm inputs $\mathsf{TSet, stag}$ and returns a list of strings, $\mathsf{t}$ (returns the data which is associated with the given keyword).
\end{description}

\subsubsection{Pseudorandom Functions (PRFs)}
In our framework, we use PRFs as the deterministic encryption to hide the search queries and tokens. It has also been used in the generation of $\mathsf{TSet}$ for confidentiality of indexes and keywords. A PRF \cite{prf} is a set of effective functions, where no efficient algorithm can differentiate between a randomly selected function from the PRF family and a random oracle (a function whose outputs are fixed completely at random), with a substantial advantage. Pseudorandom functions are essential tools in the construction of cryptographic primitives, defined as follows:\par
Let $X$ and $Y$ be sets, $F \colon \{0, 1\} ^{\lambda} \times X \rightarrow Y$ be a function, s$\stackrel{\$}{\leftarrow}$S be the operation of assigning to s an element of S chosen at random, $\operatorname{Fun}(X, Y)$ denote the set of all functions from X to Y, $\lambda$ denote the security parameter for the PRF, and
$\mathsf{negl}(\lambda)$ represent a negligible function.
We say that $F$ is a pseudorandom function (PRF) if for all efficient adversaries $\mathcal{A}$, $\operatorname{Adv}_{F, \mathcal{A}}^{\mathrm{prf}}(\lambda)=\operatorname{Pr}[\mathcal{A}^{F(K, \cdot)}(1^{\lambda})=1]-\operatorname{Pr}[\mathcal{A}^{f(\cdot)}(1^{\lambda})=1]\leq \mathsf{negl}(\lambda)$, where the probability is over the randomness of $\mathcal{A}$, $K\stackrel{\$}{\leftarrow}\{0,1\}^{\lambda}, \text { and } f \stackrel{\$}{\leftarrow} \operatorname{Fun}(X, Y)$.

\remove{
\subsubsection{Hardness Assumption}
Our construction's security relies on the hardness of the Decisional
Diffie-Hellman (DDH) problem. Let $\mathbb{G}$ be a group of prime order $p$ and generator $\mathsf{g}$. DDH problem is: given two ensembles ($\mathsf{g}$; $\mathsf{g}^a$; $\mathsf{g}^b$; $\mathsf{g}^{ab}$) and ($\mathsf{g}$; $\mathsf{g}^a$; $\mathsf{g}^b$; $\mathsf{g}^c$) where, $a$; $b$; $c~\in~  \mathbb{Z}_p$ chosen randomly, there is no probabilistic polynomial time (PPT) algorithm to distinguish the mentioned ensembles (any PPT distinguisher advantage is negligible).}

\subsubsection{Bloom Filter}
\begin{sloppypar}
A Bloom filter BF \cite{bf} is a probabilistic data structure used to test the membership of an element in a set of N elements. 
The idea is to select k independent hash functions, $\{H_i \}_{1\leq i \leq k}$. 
The Bloom filter consists of an $m$-bit binary vector, all bits set to $0$.
The bits at positions $\{H_i(n) \}_{1\leq i \leq k}$ are modified to $1$ for each element $n$ in the set, in order to set up BF for the set.
We search if $b$ has $1$'s in all positions $\{H_i(q)\}_{1\leq i \leq k}$ to verify membership of $q$ in the set and if so, we infer $q$ is in the set with a high probability. Otherwise, $q$ is not in the set with probability $1$.
There are possible false-positive matches when $q$ is not in the set, and yet the membership test returns $1$. The false-positive probability for $q$ over a uniformly random choice of $\{H_i\}_{1\leq i \leq k}$ is:
$P_{e} \leq {(1 - e^{-k \cdot N\slash m})}^k$. We use the Bloom filter to keep a part of encrypted database in a small fraction of RAM on the server.
\end{sloppypar}
\color{black}

\section{System Design and Security Model}
In this section, first, the used notations are given and then, a general architecture of PrivGenDB, and threat model is illustrated. Finally, privacy requirements and security definitions are discussed. 

\subsection{Notations}

Frequently used notations in this paper are listed in Table \ref{table:notations}.
\color{black}
\begin{table}[htb]
\caption{Notations}\label{table:notations}
\scriptsize
\begin{tabular}{>{\centering\arraybackslash} m{0.34\linewidth}  m{0.56\linewidth}}
\toprule
\textbf{Notation} & \textbf{Description} \\
\midrule
$\DOID$& Data Owner's unique $\mathsf{ID}$\\ \hline $\mathsf{ID}'_O$& Encrypted Data Owner's $\mathsf{ID}$\\ \hline $\mathcal{S}$& The set of all SNP indeices $s$\\ \hline $\Theta_s$& The set of all genotypes appeared in $\mathsf{SNP}_s$, e.g., $\Theta_3=$\{CC, TT, CT\} in Table \ref{table:Data Representation with SNPs}\\ \hline $\mathsf{G}_s=\{g=s||\theta_s\colon \theta_s\in\Theta_s\}$& The set of all keywords $g$ formed by concatanating SNP index to all genotypes of that particular SNP\\ \hline $\Phenotype$& The set of all keywords $g$ related to phenotypes\\ \hline $\Metadata$& The set of all keywords $g$ related to other information such as gender and ethnicity\\ \hline $g\in \G=\{\mathsf{G}_s\}_{s\in\mathcal{S}}\cup\Phenotype\cup\Metadata$& List of all keywords $g$ in the  database\\ \hline $\G_{\DOID}$& List of keywords $\DOID$ has, which defines the genotypes, phenotypes and other information related to a particular Data Owner\\ \hline $\DBGePh$($\g$)=\{$\DOID: \g \in \G_{\DOID}$\}& The set of Data Owner IDs that   contain that particular $\g$, which is either genotype or phenotype or   information like gender\\ \hline $\DBGePh$& Genotype-Phenotype $\mathsf{D}$ata$\mathsf{B}$ase\\ \hline $\mathsf{E}\DBGePh$& Encrypted Genotype-Phenotype $\mathsf{D}$ata$\mathsf{B}$ase\\ \hline $\varrho$& The exact predicates being asked   in a query\\ \hline $\mathsf{sterm}$& The least frequent $\varrho$ among predicates in the query\\ \hline $\mathsf{xterm}$& Other predicates ($\varrho$) in the query (except $\mathsf{sterm}$)                                                           \\
\bottomrule
\end{tabular}
\end{table}

\subsection{System Design Overview}
The proposed system is composed of different entities, as shown in Fig. \ref{fig:system design overview}. Their functions are described as follows:\par



\begin{figure}[htb]
\centering
\includegraphics[scale=0.52]{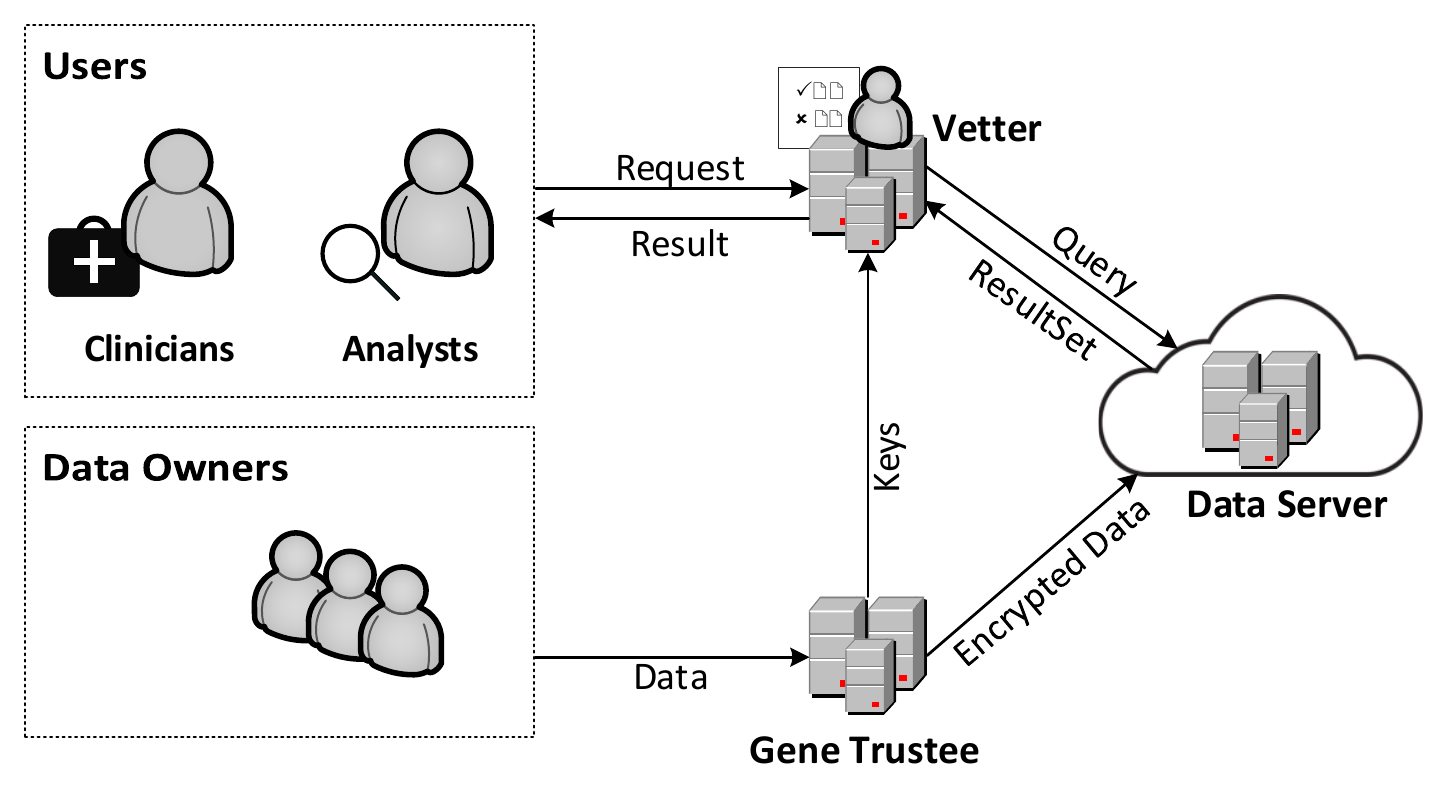}
\caption{System model overview}
\label{fig:system design overview}
\end{figure}

\remove{
\begin{figure}[htb]
\centering
\subfigure[System Model Overview]{
\includegraphics[scale=0.52]{figs/system_model_overview.pdf}
}
\subfigure[Sequence Diagram of our Proposed Model]{
\includegraphics[scale=0.58]{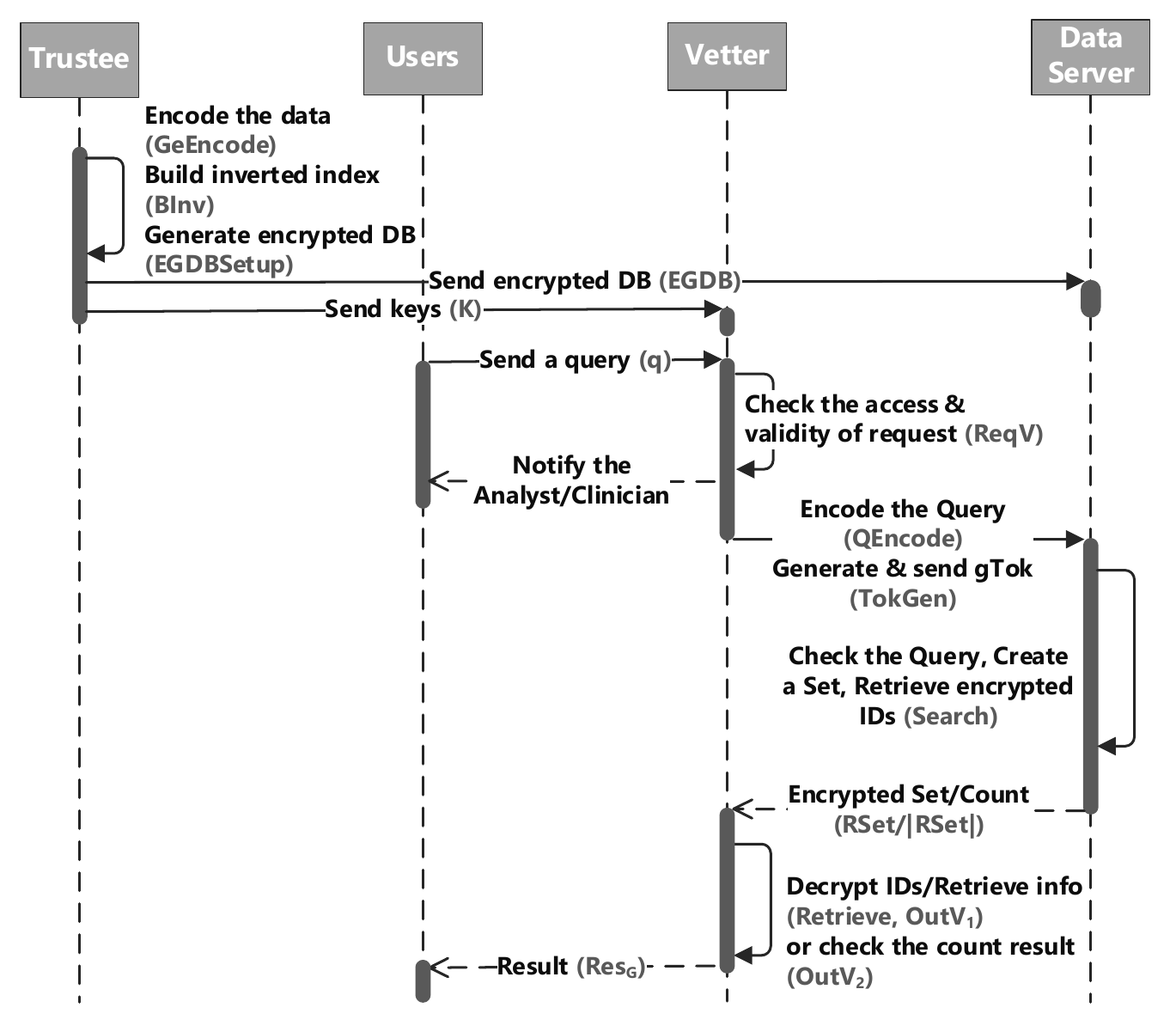}
}
\caption{System Design and Sequence Diagram of PrivGenDB}
\label{fig:system design overview}
\end{figure}
}

\begin{itemize}
\item{Data owner ($\DO$):} A data owner is a person whose data is gathered. When a data owner arrives at a medical institution like a gene trustee, as a study participant or a patient, her data is collected and recorded while her consent is given to the trustee to use her genome data for future research or treatment.
\item{Data provider or Trustee ($\Trustee$):} A medical institution, such as gene trustees, acts as a data provider in PrivGenDB. 
$\Trustee$ keeps a list of collected genome data with consent related to them. 
We assume that the data provider is trustworthy. The main responsibilities of this entity are: (i) encoding data and generating an inverted index, (ii) encrypting the generated inverted index, (iii) managing the cryptographic keys. 
\remove{\begin{itemize}
\item Encoding data and generating an inverted index.
\item Encrypting the generated inverted index.
\item Managing the cryptographic keys.
\end{itemize}}
\item{Users -- Analysts or Clinicians -- ($\User$):}
Users are either analysts who want to get the count results for analysis/research purposes, or clinicians who want to get information about the patients. They send their detailed requests to the vetter and wait for the result of the query execution. Different queries are allowed for different users (i.e. analysts may execute count queries, and clinicians may execute count/boolean/$k'$-out-of-$k$ match queries). 
\item{Vetter ($\Vetter$):} This is a trusted role to check the access privileges and vet the request submitted to the system.
Vetter $\Vetter$ approves/denies the request, which is sent by the analyst/clinician. Furthermore, there is another vetting process when the result is being released.
\item{Data Server ($\Server$):} The data server records genomic data, phenotype, and other information provided by the data provider. The $\Server$ executes the encrypted queries on encrypted data and sends back the result.
\end{itemize}



\begin{rmk}
Our ideal security goal is not to let the Data Server ($\mathcal{D}$) learn anything about the shared genomic data and the unencrypted results of the executed query by the analysts/clinicians.
The Data Server is semi-honest (honest-but-curious adversary). That means $\mathcal{D}$ correctly follows the protocol and has no intention of maliciously behaving to produce the wrong result. However, $\mathcal{D}$ may try to gain more information than what is expected to be extracted during or after the execution of the protocol. We assume the trustee to be a trusted entity. Users (Analysts or Clinicians) can be malicious, so they will be authorised through the vetter, which herself is a trusted entity checking the legitimacy of the query. We finally assume that $\mathcal{D}$ does not collude with $\User$, and $\Vetter$ receives the correct keys from $\Trustee$.
Our additional goal is to achieve fast processing time by the data server that is scalable to large databases. To reconcile these two conflicting goals, our actual protocol (see Section \ref{section:secdef}) leaks a small amount of information to the $\mathcal{D}$.
\end{rmk}

\newtheorem{definition}{Definition}

\subsection{Privacy Patterns/Requirements}\label{label:privacyrequirements}
The following privacy patterns can be defined when outsourcing genomic data and conducting different queries against it \cite{hasan2018secure}:
\begin{itemize}
    \item Data Privacy (DP): the genomic data stored in the cloud server should be protected, and no information about the data should be leaked. The confidentiality of the data should be preserved, even though the cloud server is compromised.
    \item Query Privacy (QP): entities such as the cloud server, data owners, and the compromised server should not learn anything beyond what is predicted about the query, which is executed by a researcher/clinician.
    \item Output Privacy (OP): the result of the query should not be revealed to other entities except the intended recipient.
\end{itemize}

The purposes behind requiring these privacy patterns are: (i) to protect the individuals' data from data server or any other entity trying to get their information (DP), (ii) to preserve the clinicians' patients' records or analysts' research objectives from the server (QP), and (iii) to provide privacy for individuals whose records are retrieved and privacy for analysts/clinicians' objectives (OP).

For privacy leakage through output, some privacy policies through $\Vetter$ can be considered, like defining threshold for the count query output or referring to a logbook for checking the queries already submitted to the system by a particular user not to let her conduct another particular query. These policy decisions can be made by the organisation and limit the privacy leakage, which can be gained through output results. This part is out of the scope of this research. We discuss leakages to the $\User$ in Section \ref{section:secdef}.

\subsection{Security Definitions of Our Protocol}\label{section:secdef}
Here, we discuss the security definitions for our model and its leakage profile $\mathcal{L}=\{\mathcal{L}_\mathcal{D}, \mathcal{L}_\mathcal{U}$\}. 
\subsubsection{Privacy against Data Server}\label{leakage server}
The security definition of our protocol, $\Pi_\G$, is parameterised by a leakage function $\mathcal{L}_\mathcal{D}$ (knowledge about the database and queries gained by the data server $\mathcal{D}$ through interaction with a secure scheme). This shows $\mathcal{D}$’s view in an adaptive attack (database and queries are selected by $\mathcal{D}$) can be simulated using only the output of $\mathcal{L}_\mathcal{D}$. Below, we define a real experiment $\text { Real }_{\mathcal{A}}^{\Pi}(\lambda)$ and an ideal experiment $\text { Ideal }_{\mathcal{A}, \mathcal{S}im}^{\Pi}(\lambda)$ for an adversary  $\mathcal{A}$ and a simulator $\mathcal{S}im$, respectively:
\begin{description}
\item $\text { Real }_{\mathcal{A}}^{\Pi}(\lambda)$: $\mathcal{A}(1^\lambda)$ chooses $\DBGePh$ and a list of queries $\mathsf{q}$. The experiment then runs $(\mathsf{K}$, $\mathsf{E}\DBGePh) \leftarrow \mathsf{Initialisation(\lambda, \DBGePh)}$ and gives E$\DBGePh$ to $\mathcal{A}$. Then $\mathcal{A}$ repeatedly chooses a query $\mathsf{q(\rrho)}$. The experiment runs the algorithm $\textbf{QEncode}$ on input $\mathsf{q(\rrho)}$ to get $\mathsf{q(\g)}$ and $\textbf{TokGen}$ on inputs ($\mathsf{q(\g), K)}$, and returns search tokens to $\mathcal{A}$. Eventually, $\mathcal{A}$ returns a bit that experiment uses as its output.
\item $\text { Ideal }_{\mathcal{A}, \mathcal{S}im}^{\Pi}(\lambda)$: The game starts by setting a counter $i=0$ and an empty list $\mathsf{q}$. $\mathcal{A}(1^\lambda)$ chooses a $\DBGePh$ and a query list $\mathsf{q}$. The experiment runs $\mathsf{E}\DBGePh \leftarrow \mathcal{S}im(\mathcal{L}_\mathcal{D}(\DBGePh))$ and gives $\mathsf{E}\DBGePh$ to $\mathcal{A}$. Then, $\mathcal{A}$ repeatedly chooses a search query $\mathsf{q}$. To respond, the experiment records this query as $\mathsf{q}[i]$, increments $i$ and gives the output of $\mathcal{S}im(\mathcal{L}_\mathcal{D}(\DBGePh, \mathsf{q}))$ to $\mathcal{A}$, where $\mathsf{q}$ consists of all previous queries as well as the latest query issued by $\mathcal{A}$. Eventually, the experiment outputs the bit that $\mathcal{A}$ returns.
\end{description}

\begin{definition}[Leakage to Data Server]\label{def:leakage1} We define $\mathcal{L}_\mathcal{D}(\DBGePh, \mathsf{q})$ of our scheme for $\DBGePh=\{(\DOID, \g)\}$, $r$ equals number of Data Owners
and $\mathsf{q}=(\g_1[i],\g_x[i])$, where $\g_1$ is the $\mathsf{sterm}$ and $\g_x$ is the $\mathsf{xterms}$ in the query, as a tuple $(\mathsf {N},\bar {s},\mathsf {SP},\mathsf {RP}, \mathsf {IP}, \mathsf {XP}, \mathsf {QT}, \mathsf {NP})$, where the first six components are exactly those appeared in the leakage function of \cite{cash2013highly} and the last two components are:
\begin{description}
\item  $\mathsf {QT}$ (Query Threshold): is equal to $k'$ which is the threshold given to $\mathcal{D}$ in the token so that $\mathcal{D}$ can respond to the $k'$-out-of-$k$ query.
\item  $\mathsf {NP}$ (Negation Pattern): The number of negated and non-negated predicates $\g$, which are $\mathsf{xterms}$ $\g_x$ in the query, and will be sent in two different tokens $\gtoken_1$ and $\gtoken_2$ to the data server.
\end{description}
\end{definition}

\begin{rmk}
The first six leakages mentioned in $\mathcal{L}_\mathcal{D}(\DBGePh, \mathsf{q})$ are derived from OXT leakage in \cite{cash2013highly}. The other ones are minor leakages related to the fact that SSE is being exploited in a different way in our protocol and our scheme supports other queries on the genomic dataset.
\end{rmk}
\begin{definition}[Security] The protocol $\Pi_\G$ is called $\mathcal{L}_\mathcal{D}$-semantically-secure against adaptive attacks if for all adversaries $\mathcal{A}$ there exists an efficient algorithm $\mathcal{S}im$ such that $|\Pr[\mathsf{Real}_{\mathcal{A}}^{\Pi}(\lambda) =1] - \Pr[\mathsf{Ideal}_{\mathcal{A, S}im}^{\Pi}(\lambda) =1]| \leq \mathsf{negl}(\lambda)$.
\end{definition}

\begin{algorithm*}[hbt!]
  \scriptsize
\caption{\textbf{$\mathsf{PrivGenDB.Initialisation}$}}
        \label{alg: Initialisation}
\begin{multicols}{2}
\underline{$\mathsf{PrivGenDB.Initialisation}(\lambda, \DBGePh)$}\newline
$\mathsf{Input: \lambda, \DBGePh}$\newline
$\mathsf{Output:}$ $\mathsf{E}\DBGePh, \K$\newline
The $\Trustee$rustee performs:

\begin{algorithmic}[1]
\State $\G\gets \textbf{GeEncode}(\DBGePh)$
\State $\mathsf{IINX} \gets \textbf{BInv}(\G, \DBGePh)$
\State ($\mathsf{E}\DBGePh, \mathsf{K}) \gets$ $\textbf{EGDB.Setup}(\lambda, \mathsf{IINX}, \DBGePh)$
\State $\Trustee$rustee outsources $\mathsf{E}\DBGePh$ to the $\Server$.
\State $\Trustee$rustee sends {$\K=(\K_S, \K_X, \K_I, \K_Z, \K_T)$} to the $\Vetter$etter.
  \end{algorithmic}
\underline{$\textbf{GeEncode}(\DBGePh)$}\newline
$\mathsf{Input:\DBGePh}$\newline
$\mathsf{Output:}$ $\G$
\begin{algorithmic}[1]
\State $\ell \gets$ the number of $\mathsf{SNP}$s in $\DBGePh$
\For {$\mathcal{S}=1, \ldots,\ell$}
\State $\SNP$ $\gets \{\}$
 \For {each distinct genotype $\theta_s$ in column:``$\mathsf{SNP}_{s}$" }
 \State generate $\g=s||\theta_s$
 \State $\SNP \gets \SNP\cup {\g}$
 \EndFor
\EndFor
\State $\Phenotype$ $\gets \{\}$ and $\Metadata$ $\gets \{\}$
\For {each distinct keyword $\rho$ in column:``Phenotype"}
\State $\Phenotype \gets \Phenotype\cup {\rho}$
\EndFor
\For {each distinct keyword $\Delta$ in columns:``Gender, Ethnicity"}
\State $\Metadata \gets \Metadata\cup {\Delta}$
\EndFor
\State \Return $ \G =\{\mathsf{G}_s\}_{s\in\mathcal{S}}\cup\Phenotype\cup\Metadata$.
  \end{algorithmic}
\underline{$\textbf{BInv}(\G, \DBGePh)$}\newline
$\mathsf{Input:\G, \DBGePh}$\newline
$\mathsf{Output:}$ $\mathsf{IINX}$
\begin{algorithmic}[1]
\State Initialize ${\mathsf{IINX}}$ to an empty array indexed by keywords in $\G$

\For {each keyword $\g \in \G$}
\State Output ($\g$, $\DOID$)
\EndFor

\For {each keyword $\g \in \G$}
\For{all $\DOID$ appended to $\g$}
\State $\DBGePh(\g)\leftarrow\DOID$
\EndFor
\State $\mathsf{IINX}\leftarrow$($\g$, $\DBGePh(\g)$)
\EndFor


 \State \Return $\mathsf{IINX}$   
  \end{algorithmic}
 \underline{$\textbf{EGDB.Setup}(\lambda, \mathsf{IINX})$}\newline
$\mathsf{Input:}$$\lambda, \mathsf{IINX}$\newline
$\mathsf{Output:}$ $\mathsf{E}\DBGePh, \K$
 \begin{algorithmic}[1]
\State Select key $\K_S$ for PRF $F$ and keys $\K_X,~\K_I,~\K_Z$ for PRF $F_p$ using security parameter $\lambda$ 
, and $\mathbb{G}$ a group of prime order $p$ and generator $h$.
 \State Parse $\DBGePh$ as $(\DOID, \g)$ and $\GSet$ $\gets \{\}$
\For {$\g \in {\G}$}
\State Initialize ${\bf \mathsf{inv}}\gets\{\};$ and let $\K_e \gets F(\K_S,\g)$.
\For {$\DOID\in \DBGePh(\g)$}
\State Set a counter $c\gets 1$
\State Compute $\mathsf{X}\DOID\gets F_p(\K_I,\DOID)$, $z\gets F_p(\K_Z,\g||c); y\gets \mathsf{X}\DOID. z^{-1}$,  $\DOID' \gets {E}(\K_e,\DOID)$.
\State Set $\gtag \gets h^{F_{p}(\K_{X}, g)\cdot \mathsf{X}\DOID}$ and $\GSet \gets \GSet\cup {\gtag}$
\State Append $(y,\DOID')$ to ${\bf \mathsf{inv}}$ and $c\gets c+1$.
\EndFor
\State ${\bf \mathsf{IINX}}[\g] \gets {\bf \mathsf{inv}}$
\EndFor
\State Set $(\GInv,\K_T)\gets {\mathsf{TSet.Setup}}({\bf \mathsf{IINX}})$ and
let $\mathsf{E}\DBGePh = (\GInv, \GSet)$.
\State \Return{$\mathsf{E}\DBGePh$}
 and $\K=(\K_S, \K_X, \K_I, \K_Z, \K_T)$\vspace*{-.4cm}
\end{algorithmic}
\end{multicols}
\end{algorithm*}
\begin{algorithm*}[hbt!]

  \scriptsize
\caption{\textbf{$\mathsf{PrivGenDB.QuerySubmission}$}}
\label{alg: QuerySubmission}
  \begin{multicols}{2}
\underline{$\mathsf{PrivGenDB.QuerySubmission}(\User, \mathsf{q}(\rrho_{1},...,\rrho_{n}), \K)$}\newline
$\mathsf{Input:\User, q(\rrho_{1},...,\rrho_{n})}, \K$\newline
$\mathsf{Output:}$ $\gTok, \gamma$\newline
The $\Vetter$etter performs:
\begin{algorithmic}[1]
\State ${b} \gets \textbf{ReqV}(\User, \mathsf{q}(\rrho_{1},...,\rrho_{n}))$   

\If {$b$=$1$} 
\State $\mathsf{q}(\g_{1},...,\g_{n}) \gets \textbf{QEncode}(\mathsf{q}(\rrho_{1},...,\rrho_{n}))$
\State $(\gTok, \gamma) \gets \textbf{TokGen}(\mathsf{q}(\g_{1},...,\g_{n}), \K)$
\State $\Vetter$etter sends $\gTok$ to the $\Server$
\EndIf
\State \textbf{else} reject the request and inform the $\User$
  \end{algorithmic}
\underline{$\textbf{TokGen}(\mathsf{q}(\g_{1},...,\g_{n}), \K)$}\newline
$\mathsf{Input:}\mathsf{q}(\g_{1},...,\g_{n}), \K$\newline
$\mathsf{Output:}$ $\gTok$
\begin{algorithmic}[1]
\State Computes ${\stag}\gets {\mathsf{TSet.GetTag}}(\K_T,\g_1)$.
\State $\Vetter$ sends ${\stag}$ to $\Server$.\newline
// Based on the type of the query, $\Vetter$ generates $\gTok$:\newline
---------------\textcolor{blue}{Boolean/Count}, \textcolor{purple}{$k'$-out-of-$k$ match} Query------------------
\For {$c =1, 2, \ldots$ until $\Server$ stops}
                \For {$i=2, \ldots,n$}
                  \State  $\begin{array}{rcl} \gtoken[c, i] & \gets & h^{F_p(\K_Z,\g_1||c)\cdot F_p(\K_X,\g_i)}  
                   \end{array}$            
                \EndFor
        \State \textcolor{blue}{${\gtoken}_{1}[c] \gets (\gtoken[c,2],\ldots,\gtoken[c,n])$}  \newline\textcolor{blue}{//for non-negated terms} 
\State \textcolor{blue}{${\gtoken}_{2}[c] \gets (\gtoken[c,i],\ldots)$}  \newline
                    \textcolor{blue}{//for negated terms}
\State \textcolor{purple}{$\gtoken[c] \gets (\gtoken[c,2],\ldots,\gtoken[c,n])$ //for all terms}
        \EndFor
    \textcolor{blue}{\If{Query=Boolean}
        \State $\gamma \gets 1$
       \State  {\textbf{else}}
        \State $\gamma \gets 2$
        \EndIf}
        \State \textcolor{blue}{$\gTok\gets (\stag, {\gtoken}_1, {\gtoken}_{2}, \gamma)$}
\textcolor{purple}{ \State $\gamma \gets 3$}
\State \textcolor{purple}{$\gTok\gets (\stag, \gtoken, k', \gamma)$}  \State \Return{$\gTok$}\vspace*{-.4cm} 
\end{algorithmic}
\end{multicols}
\end{algorithm*}


\subsubsection{Data Server Content Privacy against Users}
We define the leakage to the defined users in each query execution as the final output, which is vetted and is revealed to the users.

\begin{definition}[Leakage to Users] For each users' class (analysts/clinicians),  $\mathcal{L}_\mathcal{U}(\DBGePh)$ is defined as they are allowed to conduct different queries. The leakages to the analysts and clinicians are denoted by $(\mathsf{CN})$ and $(\mathsf{CN},\mathsf{DOR})$, respectively and defined as:\par

\begin{description}
\item  $\mathsf {CN}$ (Count Number): is equal to the number of data owners whose genomic/phenotypic data matches the predicates in the query.
\item  $\mathsf {DOR}$ (Data Owner Record): is the records of patients that contains some individual information of that patient when it needs to be given to the clinician for treatment purposes.
\end{description}
\end{definition}

\section{PrivGenDB: privacy-preserving query executions on genomic data}\label{construction}
\begin{figure}[htb]
\centering

\includegraphics[scale=0.58]{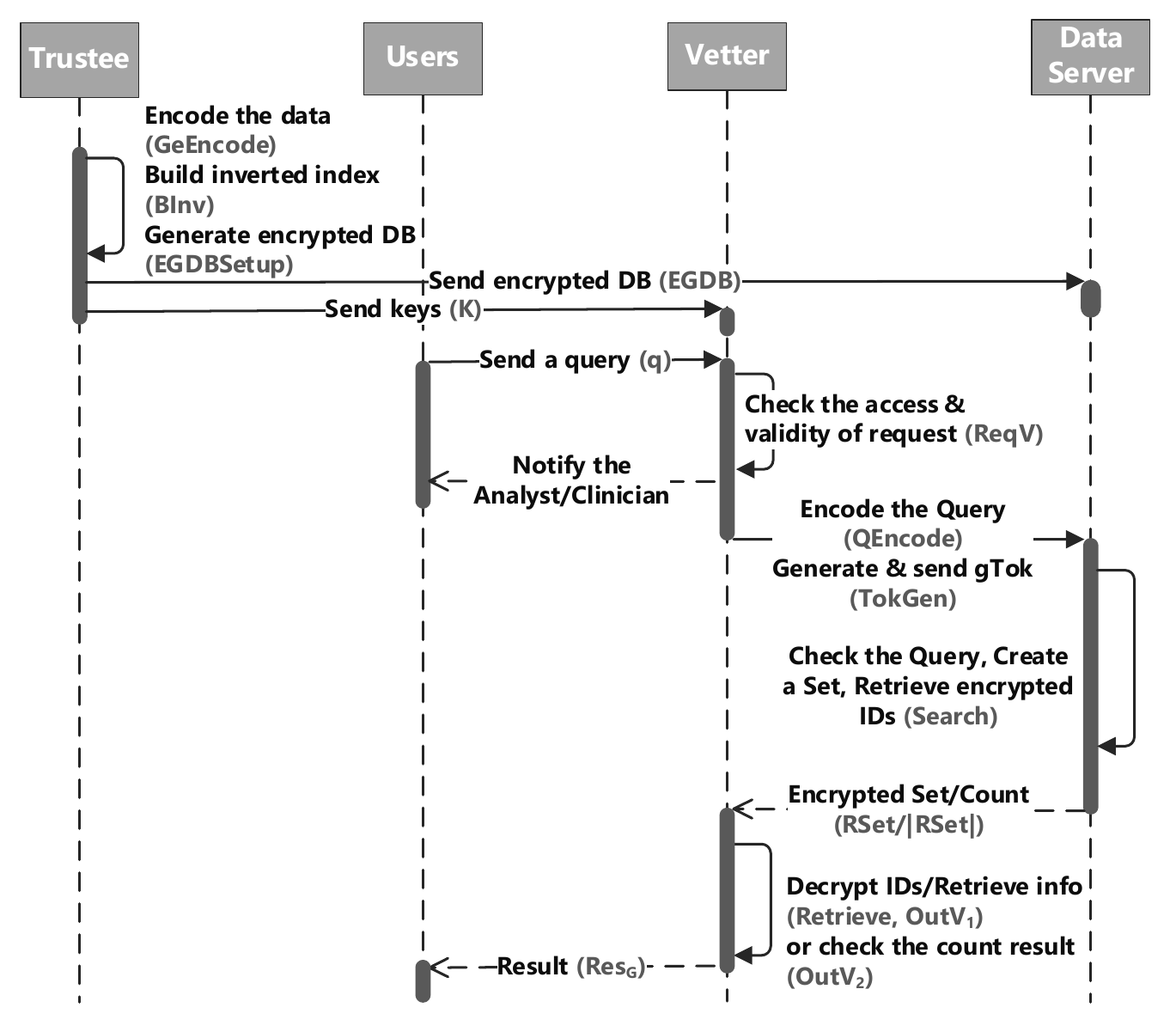}
\caption{Sequence diagram of PrivGenDB}
\label{fig:system design overview2}
\end{figure}

This section provides a detailed architecture of PrivGenDB with the designed algorithms. 
PrivGenDB consists of four phases: $\Pi_\G = (\mathsf{PrivGenDB.Initialisation}$, $\mathsf{PrivGenDB.QuerySubmission}$, $\mathsf{PrivGenDB.Search}$, $\mathsf{PrivGenDB.ResGeneration})$.
The sequence diagram of our proposed model is depicted in Fig. \ref{fig:system design overview2} which summarizes the steps of our model.

\noindent\paragraph*{$\mathsf{PrivGenDB.Initialisation(\lambda, \DBGePh)}$} 
This process is depicted in Algorithm \ref{alg: Initialisation}.
The Trustee $\Trustee$ runs this algorithm. Given security parameter $\lambda$ and a genotype-phenotype database $\mathsf{\DBGePh}$ containing genotypes $\Theta$ for each $\mathsf{SNP}_s$, 
phenotypes $\in \Phenotype$, other information $\in \Metadata$, this algorithm outputs the encrypted database $\mathsf{EGDB}$.

This algorithm uses following sub-algorithms:
\begin{itemize}[noitemsep,topsep=0pt,leftmargin=10pt]
    \item 
    $\textbf{GeEncode}(\DBGePh)$: 
    $\Trustee$ generates the keywords $\g \in \SNP$ by 
    concatenating the index of SNP to each $\theta_s \in \Theta_s$, $\g = s||\theta_s$. For example, based on Table~\ref{table:Data Representation with SNPs}, there would be three defined keywords for SNP$_1$, those are: 1AG, 1GG, 1AA. When $\Trustee$ receives the blood samples, sequences it and extracts the variations from it, creates a table like Table \ref{table:Data Representation with SNPs}.
    Apart from the genotype for each SNP, $\Trustee$ collects other information of the data owner, including phenotype, gender, and ethnicity. The $\Trustee$ also creates $\Phenotype$ for phenotype and $\Metadata$ for other columns containing information, like gender and ethnicity. For example, for Table \ref{table:Data Representation with SNPs}, we have:
$\{\mathsf{G}_s\}_{s=1}^{4}=\{$1AG, 1AA, 1GG, 2CC, 2CT, 3CC, 3CT, 3TT, 4AG, 4AA, 4GG$\}$, $\Phenotype=\{$Cancer A, Cancer B, Cancer C$\}$.

    \item  $\textbf{BInv}(\G, \DBGePh)$: $\Trustee$ runs this algorithm, which takes the keywords in $\G$ and the database $\DBGePh$ to generate an inverted index $\mathsf{IINX}$. 
    This generates the pairs of ($\g$, $\DOID$) for all the keywords $\g\in\mathsf{G}=\{\mathsf{G}_s\}_{s\in\mathcal{S}}\cup\Phenotype\cup\Metadata$.
    \item  $\textbf{EGDB.Setup}(\lambda, \mathsf{IINX}, \DBGePh$): $\Trustee$ executes this algorithm that takes the security parameter $\lambda$, generated inverted index $\mathsf{IINX}$, and the database $\DBGePh$ as inputs and outputs the encrypted database $\mathsf{EGDB}=(\GInv, \GSet)$ along with the set of keys, $\mathsf{K}$. The $\GInv$ is an array that presents equal-sized lists for each $\g$ that has been defined and created, and the $\GSet$ is a lookup table that contains elements computed from each ($\g,\DOID$) pair. For encrypting $\DOID$, which is the ID of the data owner, a key has been generated by using a PRF function to generate a key for each keyword $\g$ and use that key for all $\DOID$ having that particular $\g$.
    The $\mathsf{EGDB}$ is stored in the $\Server$, and the keys are sent to the $\Vetter$ to generate tokens for the search.
    In this algorithm, we run ${\mathsf{TSet.Setup}}({\bf \mathsf{IINX}})$, which resembles TSet Setup phase of the OXT scheme~\cite{cash2013highly}. 
    %
\end{itemize}

\noindent\paragraph*{$\mathsf{PrivGenDB.QuerySubmission(\User, \mathsf{q}(\rrho_{1},\ldots,\rrho_{n}), \mathsf{K})}$}
$\Vetter$ runs this algorithm (see Algorithm \ref{alg: QuerySubmission}), which takes the query $\mathsf{q}$, the key set $\mathsf{K}$ and user $\User$ as inputs and outputs the search token $\gTok$ by running below algorithms:\footnote{The main idea is to search for the least frequent term as the first keyword to lessen search complexity. Some steps in this process happen for all types of queries that PrivGenDB supports. $\Server$ performs each query by receiving a unique integer $\gamma$ from $\Vetter$.} 
\begin{itemize} [noitemsep,topsep=0pt,leftmargin=10pt]
    \item  $\textbf{ReqV}(\User, \mathsf{q}(\rrho_{1},\ldots,\rrho_{n}))$: $\Vetter$ authorizes the access of the $\User$ and also checks the query legitimacy. This algorithm outputs bit $b=1$ if $\User$ is allowed to execute the submitted query $\mathsf{q}$ and $b=0$ unless otherwise. PrivGenDB supports queries from researchers (who are interested in statistical analysis) and clinicians (who want to treat diseases of their patients or diagnose and prevent diseases by personalised medicine). Therefore, in this phase, $\Vetter$ checks the accessibility of researchers and clinicians and lets them execute their related queries.\footnote{
Different authentication/authorization mechanisms exist that can be utilised in this model, but discussing them is out of the scope of this work.}
    \item  $\textbf{QEncode}(\mathsf{q}(\rrho_{1},\ldots,\rrho_{n}))$: Based on the predicates in the query $q$, the $\Vetter$ generates the keywords for sending to $\Server$. The predicates related to the genotypes for SNPs will be encoded, and the new query will be generated as $\mathsf{q}(\g_{1},\ldots,\g_{n})$.
    Query encoding happens similar to the Initialisation phase. As an example, let the query be: Select count from sequences where:\par
{$\mathsf{SNP_{2}=CC}$ and $\mathsf{SNP_{4}=AG}$ and $\mathsf{diagnoses=CancerB}$.}\newline
This is encoded to:
$\g_{1}$=$\mathsf{2CC}$ and $\g_{2}$=$\mathsf{4AG}$ and $\g_{3}$=$\mathsf{CancerB}$.
    \item  $\textbf{TokGen}(\mathsf{q}(\g_{1},...,\g_{n}), \mathsf{K})$: This algorithm takes the new query predicates and the key set as inputs, so it generates the token for the search, $\gTok$. The ${\mathsf{TSet.GetTag}}(\K_T,\g_1)$ from TSet in OXT~\cite{cash2013highly} is invoked here. $\Vetter$ generates tokens as follows:\par
\begin{itemize}
    \item $\stag$ is generated based on the least frequent term, which in our database is the phenotype as $\g_1$.
    \item $\gamma$ is sent to $\Server$ to let it process the queries based on their corresponding type.
    \item For count/Boolean queries, one token is generated for non-negated requested genotypes/phenotypes/other information, and one token is generated for negated terms.
    \item For $k'$-out-of-$k$ match query, $\Vetter$ generates one token, since there are no negated terms and the process is just to check the number of matches based on some threshold ($k'$). Alongside the token and $\gamma$, $k'$ is also sent to $\Server$.
\end{itemize}
       
\end{itemize}
 



\noindent\paragraph*{$\mathsf{PrivGenDB.Search}(\gTok, \mathsf{E}\DBGePh)$} This algorithm takes the search token $\gTok$ and the encrypted database E$\DBGePh$ as inputs, then outputs the search result. Based on different tokens as the input for different queries (Boolean/Count/$k'$-out-of-$k$), the $\Server$ performs different processes and outputs the result of the query.
The $\mathsf{TSet.Retrieve}(\GInv,\stag)$ is used here, which resembles TSet in OXT scheme~\cite{cash2013highly}. The whole process is described in Algorithm \ref{alg: Search} (for Boolean/Count queries, please only follow the blue and black lines and for $k'$-out-of-$k$, please read the black and red lines.)

\noindent\paragraph*{$\mathsf{PrivGenDB.ResGeneration}(\mathsf{RSet/|RSet|,}  \g_1, \mathsf{T}, \mathsf{K})$} 
The $\Vetter$ runs this algorithm by getting the result input (based on the query and its result searched by $\Server$) and outputs $\mathsf{Res}_\G$ which will be sent to the $\User$. 
If the output of Search algorithm is a set, the two algorithms $\textbf{Retrieve}$ and $\textbf{OutV}_1$ will be executed by the $\Vetter$. If the output is a number, then the $\Vetter$ runs the $\textbf{OutV}_2$ algorithm, see Algorithm~\ref{alg: ResGeneration}.
\begin{itemize} [noitemsep,topsep=0pt,leftmargin=10pt]
    \item  $\textbf{Retrieve}(\mathsf{RSet}, \g_1, \mathsf{K})$: This algorithm takes the $\mathsf{RSet}$ and $\g_1$ as inputs and outputs the set of $\DOID$ by decrypting all the $\DOID'$ in the $\mathsf{RSet}$ and puts them in a new set $\mathsf{IDSet}$.
    \item  $\textbf{OutV}_1$($\mathsf{IDSet})$: This algorithm takes the $\mathsf{IDSet}$ as the input and retrieve information based on the executed IDs, if needed. Then, the vetter puts the result in $\mathsf{Res}_\G$ and sends it to the User.
    \item { $\textbf{OutV}_2$($|\mathsf{RSet}|, \mathsf{T})$:
    This algorithm is executed when the output of the search is the number of matched records with the predicates in the query. The input of this algorithm is $|\mathsf{RSet}|$ and $\mathsf{T}$ which is the threshold for the output.
    So, the $\Vetter$ checks the output based on the possessed threshold and put the result in $\mathsf{Res}_\G$ and sends it to the User if it satisfies the threshold. }
\end{itemize}

\begin{algorithm*}[hbt!]

  \scriptsize
\caption{\textbf{$\mathsf{PrivGenDB.Search}$}} 
\label{alg: Search}
   \begin{multicols}{2}

\underline{$\mathsf{PrivGenDB.Search}(\gTok,$ $\mathsf{E}\DBGePh)$}\newline
$\mathsf{Input:}\gTok$, $\mathsf{E}\DBGePh$\newline
$\mathsf{Output:}$ $\mathsf{RSet}$\newline
$\Server$ performs the search based on input $\gamma$
\begin{algorithmic}[1]
                \State $\mathsf{RSet}$ $\gets \{\}$
                \State $\mathsf{inv} \gets \mathsf{TSet.Retrieve}(\GInv,\stag)$
                \For {$c=1,\ldots,|\mathsf{inv}|$}
                    \State Retrieve $(\DOID',y)$ from the $c-$th tuple in $\mathsf{inv}$\newline
---------------\textcolor{blue}{Boolean/Count}, \textcolor{purple}{$k'$-out-of-$k$ match} Query------------------
\If { $\gamma$ = 1 or 2    } 
\textcolor{blue}{\If { ${\gtoken}_1[c, i]^{y}\in \GSet$ for all $i=2,\ldots,n$ in ${\gtoken}_1$ and ${\gtoken}_2[c, i]^{y}\not\in \GSet$ for all $i$ in ${\gtoken}_2$}
                            \State $\mathsf{RSet}$ $ \gets$ $\mathsf{RSet}$$\cup \{\DOID'\}$
                      \EndIf}
\State \textbf{else}
 \textcolor{purple}{\State j $\gets 0$  } 
                      \textcolor{purple}{ \For {$i=2,\ldots,n$}
                      \textcolor{purple}{  \If { $\gtoken[c, i]^{y}\in \GSet$ }
                        \State j $\gets$ j+1
                        \EndIf}
                      \textcolor{purple}{  \EndFor}
                   \textcolor{purple}{    \If {j $\geq$ $k'$}
                         \State $\mathsf{RSet} \gets$ $\mathsf{RSet}\cup \{\DOID'\}$
                        \EndIf}}
                        \EndIf
\EndFor
                      
                        \If{$\gamma = 1$ or $3$}
                        \State \Return{$\mathsf{RSet}$}
                        \State {\textbf{else}}
                        \State \Return{$\mathsf{|RSet|}$}

                \EndIf\vspace*{-.4cm}
\end{algorithmic}
 

 \end{multicols}
\end{algorithm*}
\begin{algorithm*}[hbt!]

  \scriptsize
\caption{\textbf{$\mathsf{PrivGenDB.}\mathsf{ResGeneration}$}}
\label{alg: ResGeneration}
   \begin{multicols}{2}
\underline{$\mathsf{PrivGenDB.ResGeneration}(\mathsf{RSet/|RSet|,}  \g_1, \mathsf{T,}\K)$}\newline
$\mathsf{Input:}\mathsf{RSet/|RSet|},  \g_1, \mathsf{T}, \K$\newline
$\mathsf{Output:}$ $\textsf{Res}_{\G}$\newline
The $\Vetter$ performs:
  \begin{algorithmic}[1]
\If {the output is a set $\mathsf{RSet}$}
\State $\mathsf{IDSet} \gets \textbf{Retrieve}(\mathsf{RSet}, \g_1, \mathsf{K})$
\State $\textsf{Res}_{\G} \gets$ $\textbf{OutV}_1 (\mathsf{IDSet})$
\EndIf
\State \textbf{otherwise} $\textsf{Res}_{\G} \gets$ $\textbf{OutV}_2(|\mathsf{RSet}|, \mathsf{T})$
\State $\Vetter$etter sends $\textsf{Res}_{\G}$ to the $\User$ser
  \end{algorithmic}
\underline{$\textbf{Retrieve}(\mathsf{RSet}, \g_1, \K)$}\newline
$\mathsf{Input:}$$\mathsf{RSet}, \g_1, K$\newline
$\mathsf{Output:}$ $\mathsf{IDSet}$
 \begin{algorithmic}[1]
  \State $\mathsf{IDSet}$ $\gets \{\}$
  \State $\Vetter$etter sets $\K_e \gets F(\K_S, \g_1)$
  \For {each $\DOID' \in \mathsf{RSet}$ received}
  \State Compute $\DOID \gets Dec(K_e, \DOID')$
  \State $\mathsf{IDSet} \gets$ $\mathsf{IDSet}\cup \{\DOID\}$
  \EndFor
\State \Return $\mathsf{IDSet}$
\end{algorithmic}
\underline{$\textbf{OutV}_1$($\mathsf{IDSet})$ }\newline
$\mathsf{Input:}\mathsf{IDSet}$ \newline
$\mathsf{Output:}$ $\textsf{Res}_{\G}$
 \begin{algorithmic}[1]
\State //$\Vetter$etter can retrieve other information if needed based on the possessed decrypted $\mathsf{IDSet}$ and put it in $\textsf{Res}_{\G}$.
\State $\Vetter$etter checks the $\textsf{Res}_{\G}$ and the $\User$.
\State \Return $\textsf{Res}_{\G}$
\end{algorithmic}
\underline{$\textbf{OutV}_2$($|\mathsf{RSet}|, \mathsf{T})$ }\newline
$\mathsf{Input:}|\mathsf{RSet}|, \mathsf{T}$ \newline
$\mathsf{Output:}$ $\textsf{Res}_{\G}$
 \begin{algorithmic}[1]
\State //$\Vetter$etter checks the threshold $\mathsf{T}$ for the output.
\If{$|\mathsf{RSet}| \geq \mathsf{T}$}
\State \Return $\textsf{Res}_{\G}$
\EndIf
\State \textbf{otherwise} the query will be rejected and the $\mathcal{U}$ser will be notified.\vspace*{-.4cm}
\end{algorithmic}

 \end{multicols}
\end{algorithm*}

\section{Security Analysis}
In this section, we prove the security of our protocol against honest-but-curious data server and the compromised users.
\newtheorem{theorem}{Theorem}
\begin{theorem} Let $\mathcal{L}_\mathcal{D}$ be the leakage function defined in Section \ref{leakage server}. Then, our protocol is $\mathcal{L}_\mathcal{D}$-semantically-secure against adaptive data server, if OXT \cite{cash2013highly} is secure.
\end{theorem}

\remove{
\begin{figure}[!b]
\centering
\includegraphics[scale=1.3]{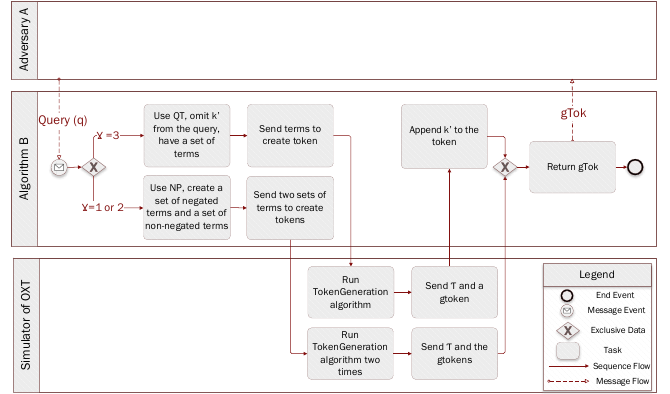}

\caption{Security reduction process}
\label{fig:security diagram}
\end{figure}
}

\begin{pf}
Let $\mathcal{A}$ be an honest-but-curious data server who performs an adaptive attack against our protocol. Then we can construct an algorithm $\mathcal{B}$ that breaks the server privacy of OXT protocol \cite{cash2013highly} by running $\mathcal{A}$ as a subroutine with non-negligible
probability. Algorithm $\mathcal{B}$ runs GeEncode on the selected $\DBGePh$ by $\mathcal{A}$ and passes the output ($\mathsf{G}$) to the OXT challenger. The OXT challenger runs (K,$\mathsf{E}\DB)\leftarrow$   OXT.EDBSetup($\mathsf{G}$) and returns $\mathsf{E}\DB$ to the algorithm $\mathcal{B}$. Then, the algorithm $\mathcal{B}$ sets $\mathsf{E}\DBGePh$ = $\mathsf{E}\DB$.
The algorithm $\mathcal{B}$ sends $\mathsf{E}\DBGePh$ to an adversary $\mathcal{A}$.
For each query $\mathsf{q}[i]$ issued by the adversary $\mathcal{A}$, algorithm $\mathcal{B}$ first runs QEncode to generate the predicates for computing tokens. Then, algorithm $\mathcal{B}$ defines TokGen(K; $\mathsf{q}[i]$) which uses the output of the TokenGeneration oracle of OXT ($\mathsf{q}[i]$ in TokGen has the encoded predicates).  
For count and Boolean queries, it categorises negated/non-negated terms and runs the TokenGeneration algorithm of OXT twice to generate $\gTok$. For $k'$-out-of-$k$ matches, it just omits the $k'$ from query sent by $\mathcal{A}$, passes the rest to TokenGeneration algorithm of OXT and generates the $\gTok$ by using that output and including $k'$ to send it to $\mathcal{A}$.
Finally, the adversary $\mathcal{A}$ outputs a bit b that the algorithm $\mathcal{B}$ returns (see Fig. \ref{fig:security diagram} for detailed security reduction process diagram-TokGen is a subprocess in this diagram which is explained above).
Since the core construction of our scheme is exploited from OXT in \cite{cash2013highly}, we can use the oracle of OXT to reduce our proof of theorem to OXT protocol. Thus, if the security of OXT holds, the security of our scheme is guaranteed.
\begin{figure}[htb]
\centering
\includegraphics[scale=1.22]{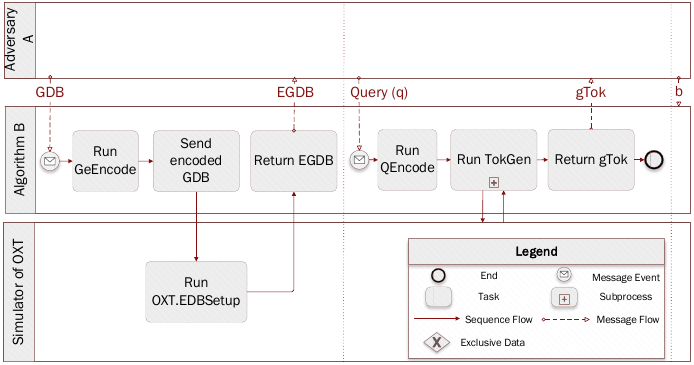}
\caption{Detailed security reduction process diagram}
\label{fig:security diagram}
\end{figure}

\remove{
\begin{figure*}[!t]
\centering
\subfigure[Security reduction process]{
\includegraphics[scale=1.23]{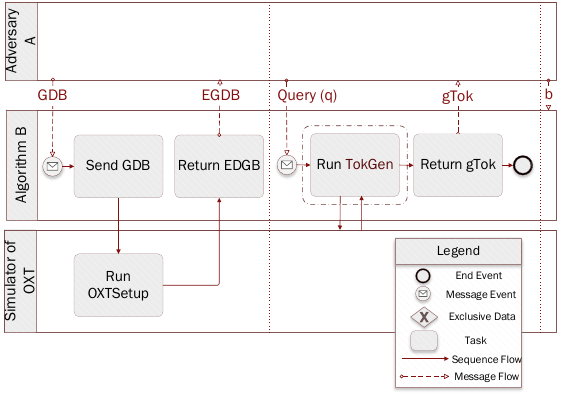}
}
\subfigure[Detailed TokGen process in the security proof]{
\includegraphics[scale=1.23]{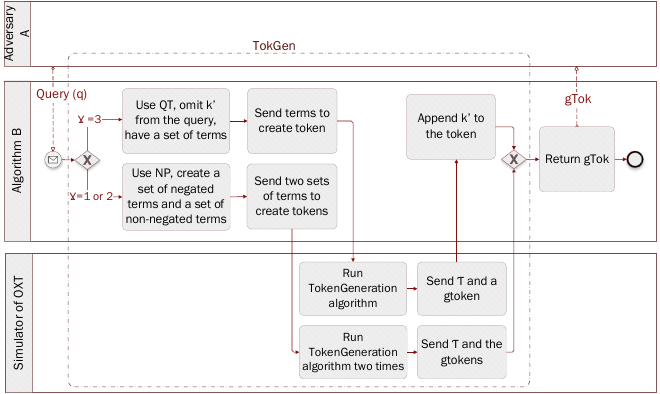}
}
\caption{Security reduction process diagram}
\label{fig:security diagram}
\end{figure*}

}

\begin{simulator}
It is only remained to present an efficient simulator $\mathcal{S}_{\G}$ that takes as input the leakage function $\mathcal{L}_\mathcal{D}$ described in Definition~\ref{def:leakage1}
and outputs an $\mathsf{EGDB}$ and simulates the tokens for the queries as follows. $\mathcal{L}_{\Pi_\mathsf{G}}$($\mathsf{GDB, q}$) = $\mathcal{L}_{\Pi_\mathsf{OXT}}$($\mathsf{GDB, q'}$), where $\mathsf{q'}$ = T($\mathsf{q}$), and
T($\mathsf{q}$) maps the query vector $\mathsf{q}$ of $\Pi_\G$ given by attacker $\mathcal{A}$, to query vector T($\mathsf{q}$) for $\mathsf{OXT}$ used by attacker $\mathcal{B}$. If $\mathsf{q}$ is a Boolean or a count query, T($\mathsf{q}$) generates two sets for negated and non-negated encoded terms in $\mathsf{q}$; hence, $\mathsf{q'}$ is two separate queries to the simulator of OXT. For $k'$-out-of-$k$ match queries, T($\mathsf{q}$) is the encoded $\mathsf{q}$ without $k'$. For these queries, such a simulator can be constructed by using $\mathcal{S}_{\mathsf{OXT}}$, a simulator for OXT protocol. 
Using all leakage components except $\mathsf{NP, QT,}$ the $\mathcal{S}_{\G}$ can simulate the $\stag, \gtoken$ by running $\mathcal{S}_{\mathsf{OXT}}$. The extra leakage components $\mathsf{NP, QT}$ are exploited by algorithm $\mathcal{B}$ to construct $\mathsf{gToK}$ completely and send it to adversary $\mathcal{A}$. 
We just need to use the simulator of $\mathsf{OXT}$ for $\mathcal{A}_\mathsf{OXT}$, to construct the simulator of our scheme for $\mathcal{A}_{\G}$. By running $\mathcal{S}_\mathsf{OXT}$, we can construct a simulator $\mathcal{S}_{\G}$ for EGDB.Setup and TokGen queries of our scheme.
\begin{flushleft}
$\operatorname{Pr}\left(\text {Real}_{\mathcal{A}}^{\Pi_{\mathrm{\G}}}=1\right)-\operatorname{Pr}\left(\text {Ideal}_{\mathcal{A}, \mathcal{S}_{\G}}^{\Pi_{\mathrm{\G}}}=1\right)\leq$ 
$\operatorname{Pr}\left(\text {Real}_{\mathcal{B}}^{\Pi_{\mathsf{OXT}}}=1\right)-\operatorname{Pr}\left(\text{Ideal}_{\mathcal{B}, \mathcal{S}_\mathsf{O X T}}^{\Pi_{\mathsf{OXT}}}=1\right)\leq \operatorname{negl}$
\end{flushleft}
Therefore, since $\mathsf{OXT}$ is secure and its advantage is negligible, our protocol's advantage is negligible.

\end{simulator}
\end{pf}
\section{Analytical Performance Comparison}
This section presents the performance comparison of our PrivGenDB with existing related works from different perspectives.  
The overall comparison is presented in Table \ref{table:comm/comp complexity}.

\begin{table*}[hbt!]
\caption{Computational and communication cost between client and server}
\label{table:comm/comp complexity}
\resizebox{\textwidth}{!}{%
\centering
\scriptsize
\begin{tabular}{cc>{\centering\arraybackslash}m{2cm}>{\centering\arraybackslash}m{2cm}>{\centering\arraybackslash}m{2cm}>{\centering\arraybackslash}m{2cm}>{\centering\arraybackslash}m{2cm}>{\centering\arraybackslash}m{2.5cm}>{\centering\arraybackslash}m{2.5cm}}
\toprule
\multicolumn{2}{c}{\textbf{Reference} }      & \begin{tabular}[c]{@{}c@{}}PrivGenDB  \end{tabular}                           & \begin{tabular}[c]{@{}c@{}}\cite{kantarcioglu2008cryptographic}\end{tabular} & \begin{tabular}[c]{@{}c@{}}\cite{canim2011secure}\end{tabular} & \begin{tabular}[c]{@{}c@{}}\cite{ghasemi2016private}\end{tabular} & \begin{tabular}[c]{@{}c@{}}\cite{nassar2017securing}\end{tabular} & \begin{tabular}[c]{@{}c@{}}\cite{hasan2018secure}\end{tabular}&
\begin{tabular}[c]{@{}c@{}}\cite{new2020}\end{tabular}\\  
\cmidrule(lr){1-2}\cmidrule(lr){3-3} \cmidrule(lr){4-4} \cmidrule(lr){5-5}\cmidrule(lr){6-6}\cmidrule(lr){7-7}\cmidrule(lr){8-8}\cmidrule(lr){9-9}
\cmidrule{1-2}\cmidrule(lr){3-3} \cmidrule(lr){4-4} \cmidrule(lr){5-5}\cmidrule(lr){6-6}\cmidrule(lr){7-7}\cmidrule(lr){8-8}\cmidrule(lr){9-9}

\multicolumn{1}{c|}{}                                 & Initialisation                 & $rx(T_E+T_F+kT_h)     + rxT_e$                                            & $rxT_{E_{pk}}$                                                                                                  & $r(\sim4x/b')T_E$                                                                    & $2rT_{E_{pk}} + $ permutation                                                                               & $4rxT_{E_{pk}}$                                                                                            & $(3+\cdots+3^x) * (2T_{E_{pk}}+kT_h)$                                                                                       & $(3+\cdots+3^x) * (4T_{E_{pk}}+kT_h)$\\ \cline{2-9} 
\multicolumn{1}{c|}{}                                 & Search (Data Server)       & $\alpha((n-1)$ $(kT_h+T_e))$                                                  & $r(n*T_M+2T_e+AC)$                                                                                          & find $n$ positions in $r$                                                                 & $O(2r) +$ \textless{}$2|Res| * T_M$                                                                      & $r(n*T_M+T_e+AC)$                                                                                     & \begin{tabular}[c]{@{}c@{}}\textgreater{}$T_M*3^{(\ell_{\mathsf{SNP}})}$       $+$GC\end{tabular}              & \begin{tabular}[c]{@{}c@{}}\textgreater{}$T_M*3^{(\ell_{\mathsf{SNP}})}$       $+$GC\end{tabular} \\ \cline{2-9} 
\multicolumn{1}{c|}{\multirow{-4}{*}{\textbf{Comp.}}} & Search\footnotemark (Client\footnotemark)       & $T_F     +\alpha(nT_F+(n-1)T_e)+T_D$                                            & $rT_{D_{pk}}$                                                                                                     & \begin{tabular}[c]{@{}c@{}}SCP:\\      $rnT_D$\end{tabular}                              & $T_{D_{pk}}$                                                                                                 & $rT_{D_{pk}}$                                                                                               & \textgreater{}$T_{D_{pk}}*3^{(\ell_{\mathsf{SNP}})}$                                                                 &\textgreater{}$T_{D_{pk}}*3^{(\ell_{\mathsf{SNP}})}$  \\ \hline
\multicolumn{1}{c|}{\textbf{Stor.}}                   & Storage Size          & $rx(\ell_E+\ell_P)+m$                                                        & $r(4xb)$                                                                                                  & $rx\ell_E$                                                                                  & $2rx+2rb$                                                                                      & $r(8xb)$                                                                                            & $(3+\cdots+3^x)*(m+2\ell_E)$                                                                                             & $(3+\cdots+3^x)*(m+4\ell_E)$\\ \hline
\multicolumn{1}{c|}{\textbf{Comm.}}                   & Bandwidth             & \begin{tabular}[c]{@{}c@{}}$\ell_F+|t|$\\      $+\alpha((n-1)\ell_D)$\end{tabular} & $r *$ “the length   of the encrypted result”                                                              & $r(b'n)$ Between   SRV and SCP                                                        & $2$len$($query$)$                                                                                        & $4$len$($query$)$                                                                                        & “the   No. of nodes need to be accessed” $* \ell_E$                                                        &“the   No. of nodes need to be accessed” $* \ell_E$ \\ \bottomrule

\multicolumn{9}{p{21cm}}{\footnotesize{$^1$ The client performs computations to generate the query or help Data Server during the search process. This computation is the token generation phase in PrivGenDB. \newline
$^2$ Client is the entity that submits the query to the server and gets back the result. It has direct contact with the server. In previous works, it is either client or researcher or a trusted entity/proxy, and in our model, it is the vetter.\newline
\textbf{Notations}: $T_e$: Time taken to compute an exponentiation; $T_F$: Time taken to compute a PRF; $T_h$: Time taken to compute a hash; $T_E$: Time taken to encrypt a block with a symmetric cryptosystem; $T_{E_{pk}}$: Time taken to encrypt a block with an asymmetric cryptosystem; $T_M$: Time taken to perform modular multiplication; $AC$: Adding constant to ciphertext; $T_D$: Time taken to decrypt a block with a symmetric cryptosystem; $T_{D_{pk}}$: Time taken to decrypt a block with an asymmetric cryptosystem;
$\alpha$: Number of records satisfying $\mathsf{sterm}$; $x$: Number of columns in the original database; $n$: Query size; $r$: Number of records in DB; $b$: Public key modulus size in bits; $b'$: AES block size; $\ell_p$:	Size of an element from $\mathbb{Z}_p$ ($p$ is a prime number);
$\ell_D$: Size of an element from Diffie-Hellman (DH) group; $\ell_F$: Size of the output of a PRF; $\ell_E$: Size of the block of SE; $\ell_h$: Size of the output of hash function H; $m$: Bloom Filter (BF) size 
GC: Garbled  circuit computation; SCP: Secure  Cryptographic  Coprocessor computation; $\ell_{\mathsf{SNP}}$:
Least requested SNP index; $|Res|$: Total number of records that matched all the predicates in the query; len(query): Length of the submitted query.}}
\end{tabular}}
\end{table*}

\begin{rmk}
In terms of functionality, our model is independent of the authentication method can be used in PrivGenDB. Therefore, to calculate the computational cost of $\mathsf{PrivGenDB.Initialisation}$, building the inverted index, and encrypting the database are considered. 
The token generation of $\mathsf{PrivGenDB.QuerySubmission}$ phase in our model has been considered as the Search (Vetter) complexity, and the whole $\mathsf{PrivGenDB.Search}$ phase of our model has been considered as Search (Data Server) complexity. 
\end{rmk}


\subsection{Initialisation Computational Cost}
In the initialisation/setup phase of all the schemes in Table \ref{table:comm/comp complexity} except \cite{hasan2018secure}, the computational cost is in the order of the number of records ($r$) multiplied by the number of columns in the original database, $x$  (which is the number of SNPs $+~1$ for phenotype $+$ the number of other characteristics if they support, e.g., gender, ethnicity) \cite{kantarcioglu2008cryptographic,canim2011secure,ghasemi2016private,nassar2017securing}. 
The tree structure proposed in \cite{hasan2018secure} needs the encryption of all nodes in the setup phase. The number of nodes increases by the number of SNPs. As acknowledged by the author of \cite{hasan2018secure}, in our correspondence, if the number of records in the database increases, the number of nodes increases as well. We also assume that each SNP can have three possibilities of genotypes, and when the number of records increases, the possibility of having a new genotype for the SNPs increases. This assumption is completely dependent on the genotypes distribution in the dataset.
The computational complexity of PrivGenDB in this phase is in the order of $rx$.
All methods in Table \ref{table:comm/comp complexity} are linear in $rx$, except \cite{ghasemi2016private} which does not encrypt all the data and \cite{hasan2018secure} which depends on the number of nodes (increases by $x$ and $r$). The proposed methods in \cite{new2018,new2020} have the same computation/communication complexities' behavior as that of \cite{hasan2018secure}, since they proposed their scheme based on the index tree in \cite{hasan2018secure}, with some added functionality or improvements.
\subsection{Search Computational Cost}
The computational costs are split between the client/trusted-entity/vetter and the server during the search process. In PrivGenDB, the vetter represents the client, as it is the entity that interacts with the server to perform the search against the database.
As it is obvious from the Table \ref{table:comm/comp complexity}, the computational cost of the search through server is in the order of the number of records in \cite{kantarcioglu2008cryptographic,canim2011secure,ghasemi2016private,nassar2017securing}. 
Based on the index tree proposed in \cite{hasan2018secure}, their scheme's search cost depends mostly on the number of SNPs. However, increasing the number of records also increases the number of nodes in their index tree to cover all the genotype possibilities (it is discussed in part A as well).
PrivGenDB's search complexity depends on the number of IDs that match the first predicate in the query ($\alpha$). 
As the phenotype was selected as the first predicate, which is the least frequent term in our database, PrivGenDB's search complexity is sublinear to the number of records in the database. 
This implies that, even by increasing the number of records, if the number of matched records to the first predicate (or phenotype) does not change, the search complexity remains unchanged.

\subsection{Storage Size}
We now investigate the storage size of PrivGenDB and compare it to other schemes. PrivGenDB stores $\GInv$ and the Bloom filter of $\GSet$ in $\mathsf{EGDB}$, while \cite{kantarcioglu2008cryptographic,canim2011secure,ghasemi2016private,nassar2017securing} store the encrypted data and \cite{hasan2018secure} stores the encrypted tree in the database.
The size of $\GInv$ equals $rx(\ell_E+\ell_P$), which is the number of pairs ($(y,\DOID')$ in the construction), multiplied by the size of each pair.
The size of Bloom filter $m$ is approximately $1.44\cdot krx$ to attain a negligible probability of false-positives.
Other schemes' storage size also depends on $rx$, except \cite{hasan2018secure}, which only supports the count query and stores the counts in the tree and depends on the number of nodes.

\subsection{Interaction Rounds and Bandwidth}
In PrivGenDB, the first interaction between the $\Vetter$ and the $\Server$ happens when the $\Vetter$ sends the search tokens to the $\Server$. Based on the type of the submitted query and its size, the size of the result sent to the $\Vetter$ differs, which defines the next interaction. Therefore, when the $\Vetter$ generates the $\stag$, sends it to the $\Server$. Then, depending on the number of retrieved IDs for the $\stag$, the $\Vetter$ generates the $\gtoken$ and sends them to $\Server$, which checks a set membership against $\GSet$.
Communication overhead in \cite{hasan2018secure} depends on the number of nodes that need to be accessed during the search to perform the query.

\section{Experimental Results}
This section presents the implementation results of PrivGenDB with a different number of records/SNPs/query-sizes as well as different types of queries. The source code is written in Java programming language, and our machine used to run the code is Intel(R) Core(TM) i7-8850H CPU @ 2.60GHz processors with 32 GB RAM, running Ubuntu Linux 18.04.
We consider the following aspects in order to assess the efficiency of our proposed method:
\begin{itemize}
    \item Initialisation Time: $\Trustee$ to perform $\textbf{BInv}$ and $\textbf{EGDB.Setup}$.
    \item Query Execution Time: $\Vetter$ to execute $\textbf{TokGen}$ and  $\Server$ to run $\mathsf{PrivGenDB.Search}$ 
    \item Storage Analysis: Storage space needed to store the database. 
    \item Communication Overhead: The data transferred between the $\Vetter$ and the $\Server$ to perform the query.
\end{itemize}
To improve the runtime performance of our implementation, we leverage an in-memory key-value storage Redis \cite{Redis} to keep the generated $\GInv$ for querying purposes. In addition, we deploy the Bloom filter from Alexandr Nikitin \cite{BF1} as it is the fastest
Bloom filter implementation for JVM. The false-positive rate of the bloom filter is set to $10 ^{-6}$, which enables our implementation to keep the $\GSet$ in a small fraction of RAM on the server.
We used the Java Pairing-Based Cryptography Library (JPBC) \cite{JPBC}.
We use both real-life and synthetic datasets for evaluation
of PrivGenDB. The real-life data are taken from The Harvard Personal Genome Project (PGP) \cite{PGP}, where we took data of $58$ patients with $2,000$ SNPs, their phenotype, gender, and ethnicity. These patients have had different types of cancer such as breast, brain, thyroid, uterine, kidney, melanoma, colon, prostate as the phenotype, or they were Covid-19 positive/negative recently. Then, we created several synthetic datasets based on the above mentioned real dataset to evaluate PrivGenDB with a different number of records ($500-40,000$) and different number of SNPs ($500-2,000$). 
\subsection{Initialisation Time}
\begin{figure}[htb]
\centering
\subfigure[\#Records=$5,000$]{
\includegraphics[scale=0.54]{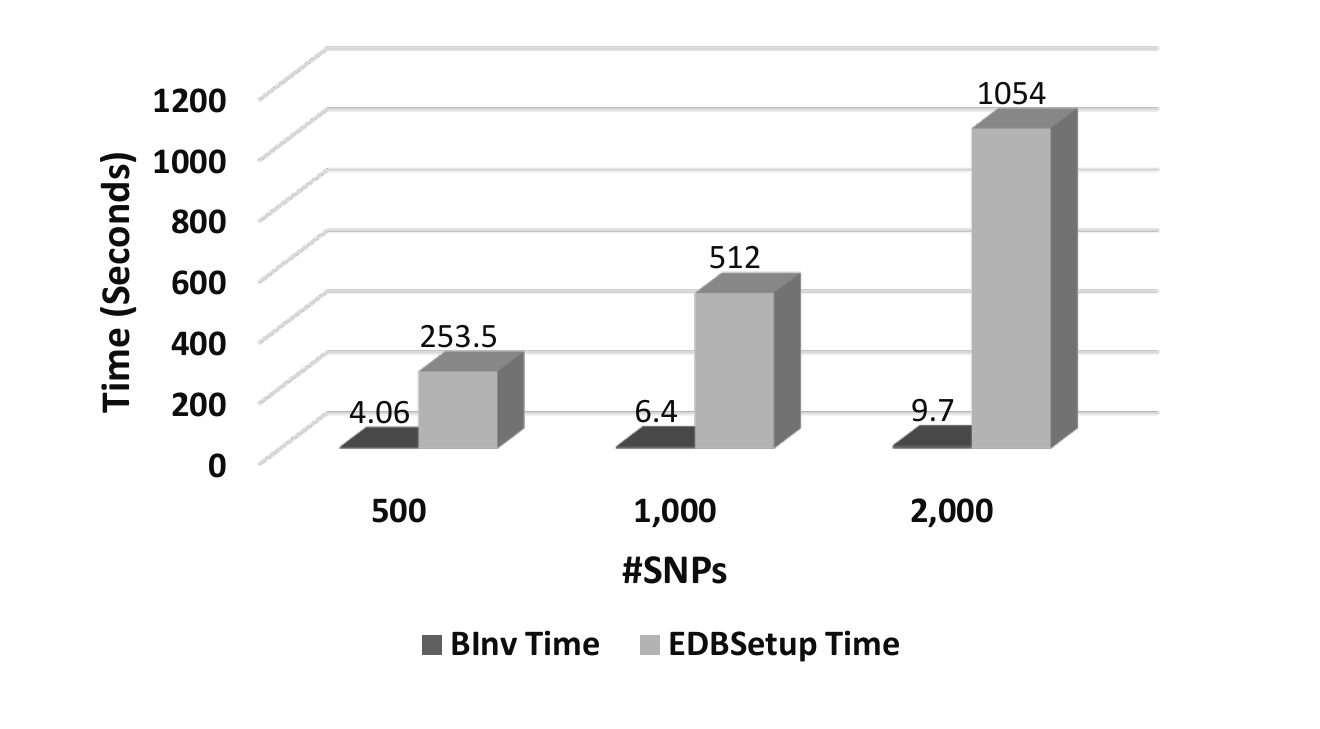}
}
\subfigure[\#SNPs=$1,000$]{
\includegraphics[scale=0.54]{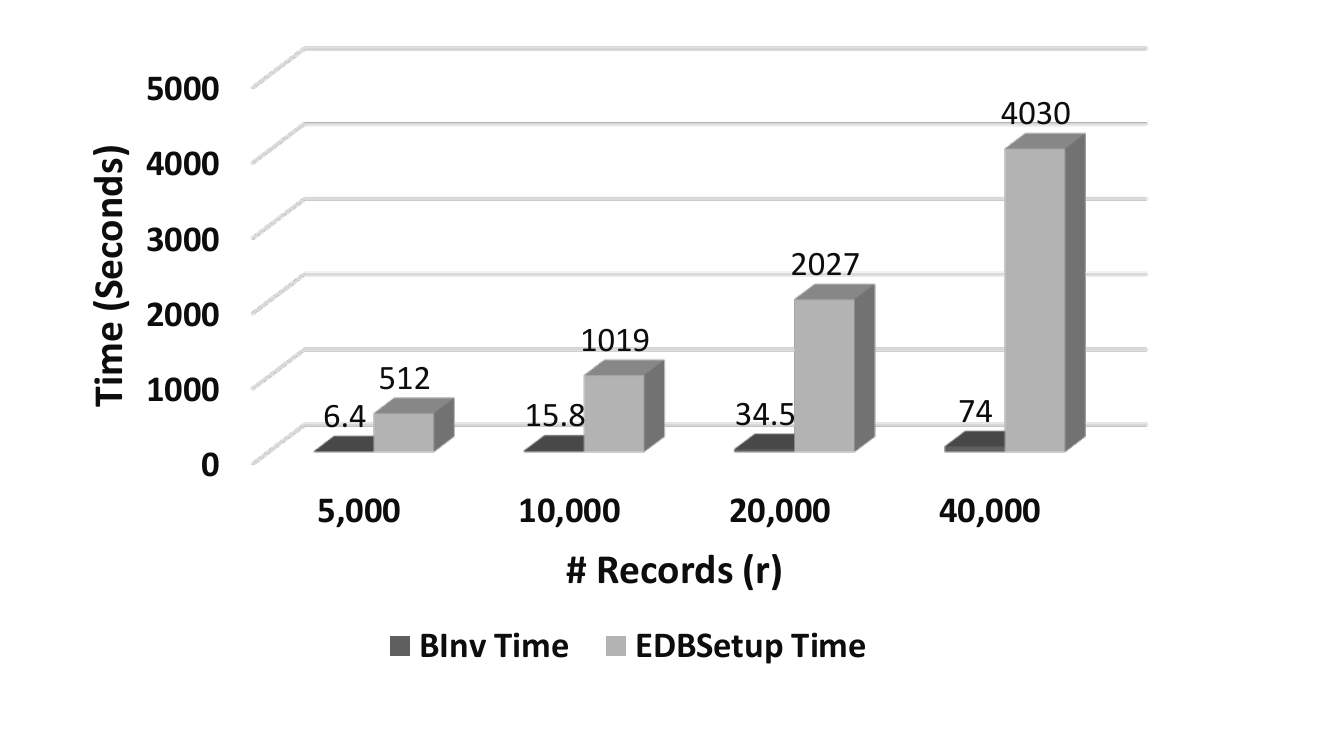}
}
\caption{Initialisation time for datasets with different number of SNPs and records. (\#SNPs=$x$-3, \#Records=$r$ in Table \ref{table:comm/comp complexity}).}
\label{fig:initialisation}
\end{figure}

Initialisation time includes the time needed for generating the inverted index and encrypting it. Genomic data encoding time is negligible. Fig. \ref{fig:initialisation} illustrates the initialisation time of PrivGenDB when the number of SNPs increases or the number of records ($r$) in the dataset increases. The time consumption for the generation of inverted index and encrypting it with $5,000$ records, $1,000$ SNPs, different phenotypes, gender and ethnicity is around $518$ seconds in PrivGenDB, while it is $47$ min and $257$ min in \cite{hasan2018secure} and \cite{ghasemi2016private}, respectively. Note that we encrypt all the information in our database to support different queries and this phase is needed to be performed once only.

\subsection{Storage Analysis}
Since PrivGenDB needs to support different queries and have the capability of retrieving the IDs, it encrypts all the related information and not just the count of different SNPs. 
Table \ref{table:Storage cost} lists the amount of spaces required to store the original data and encrypted data. The left column represents the number of records in the dataset ranging from $500$ to $40,000$. The original dataset of size $1.5$MB is encrypted to $113.4$MB, while it is $188$MB in \cite{hasan2018secure}. The expansion in the encrypted tree size of \cite{hasan2018secure} is due to encrypting the data using the Paillier Encryption.
It is worth mentioning the growth ratio in the storage overhead (expansion factor) of PrivGenDB is around $80$, while it is $180$ in \cite{hasan2018secure}, and it is in the order of $1024$ (key bit length) in \cite{kantarcioglu2008cryptographic,nassar2017securing}. 

\begin{table}
\caption{Size of original and encrypted database}
\label{table:Storage cost}
\centering
\resizebox{\columnwidth}{!}{
\begin{tabular}{ccccccc}
\toprule
\multirow{2}{*}{\#Records(r)}   & \multicolumn{2}{c}{\#SNPs=$500$}       & \multicolumn{2}{c}{\#SNPs=$1,000$}      & \multicolumn{2}{c}{\#SNPs=$2,000$}      \\ \cmidrule(lr){2-3}\cmidrule(lr){4-5}\cmidrule(lr){6-7}
 & Original  & Encrypted  & Original  & Encrypted & Original & Encrypted \\ \midrule
$500$       & $1.5$ MB         & $113.4$ MB        & $3.1$ MB       & $235$ MB        & $7.6$ MB       & $474$ MB        \\[1pt] \hline
$1,500$       & $4.4$ MB         & $336$ MB        & $8.8$ MB       & $670$ MB        & $20.2$ MB       & $1.34$ GB        \\[1pt] \hline
$5,000$       & $14$ MB         & $1.12$ GB        & $28.9$ MB       & $2.22$ GB        & $64.6$ MB       & $4.44$ GB        \\[1pt] \hline
$10,000$      & $28.5$ MB       & $2.27$ GB        & $57.8$ MB       & $4.5$ GB         & $126.1$ MB      & $8.68$ GB        \\[1pt] \hline
$20,000$      & $57$ MB         & $4.43$ GB        & $115.6$ MB      & $8.89$ GB        & $258.4$ MB      & $17.77$ GB       \\[1pt] \hline
$40,000$      & $114 $ MB      & $8.92$ GB        & $224$ MB        & $17.8$ GB       & $517$ MB        & $35.6$ GB       \\[1pt] \bottomrule
\end{tabular}}
\end{table}

\subsection{Query Execution Time}
To calculate the query execution time, we executed a number of queries with different sizes on the encrypted database.
The queries we used were determined by randomly selecting $10$, $20$, $30$, and $40$ SNP sequences with other information like gender and ethnicity. PrivGenDB works with sublinear search capability, making it a proper solution for large databases. Below, we discuss our results and show our model's scalability with respect to different characteristics of our database. Retrieve time is negligible.

\subsubsection{Scalibility of PrivGenDB in terms of the number of SNPs}
\begin{figure}
\centering
\includegraphics[scale=0.52]{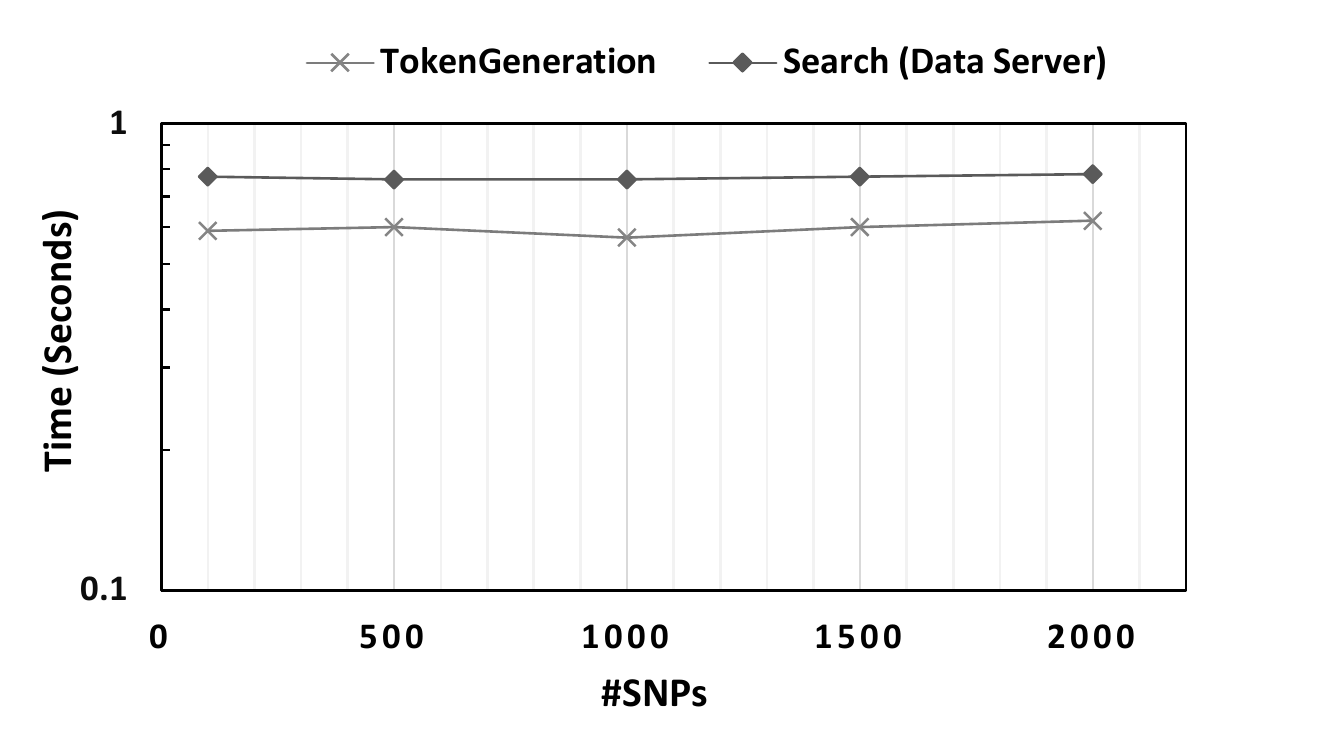}
\caption{Query execution time on datasets with $r=5,000$ and different number of SNPs, $n=10$.}
\label{fig:SearchSNP}
\end{figure}

We first experimented the effect of the number of SNPs on the query execution time. As it is obvious from Fig. \ref{fig:SearchSNP}, the time needed to perform a particular query on a dataset with $5,000$ records does not change when the number of stored SNPs for each record increases.

\subsubsection{Scalability of PrivGenDB in terms of the number of records}
\begin{figure*}[htb]
\centering
\subfigure[Token Generation]{
\includegraphics[scale=0.52]{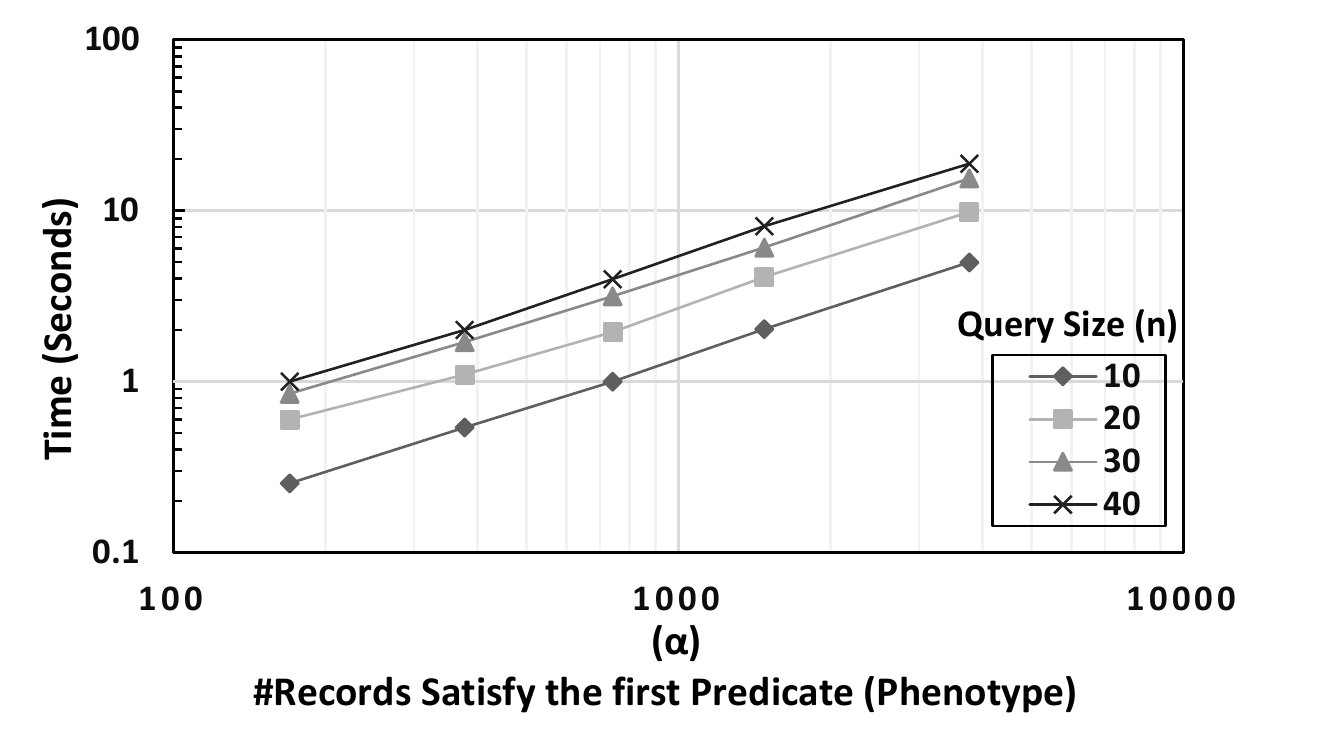}
}
\subfigure[Search (Data Server)]{
\includegraphics[scale=0.52]{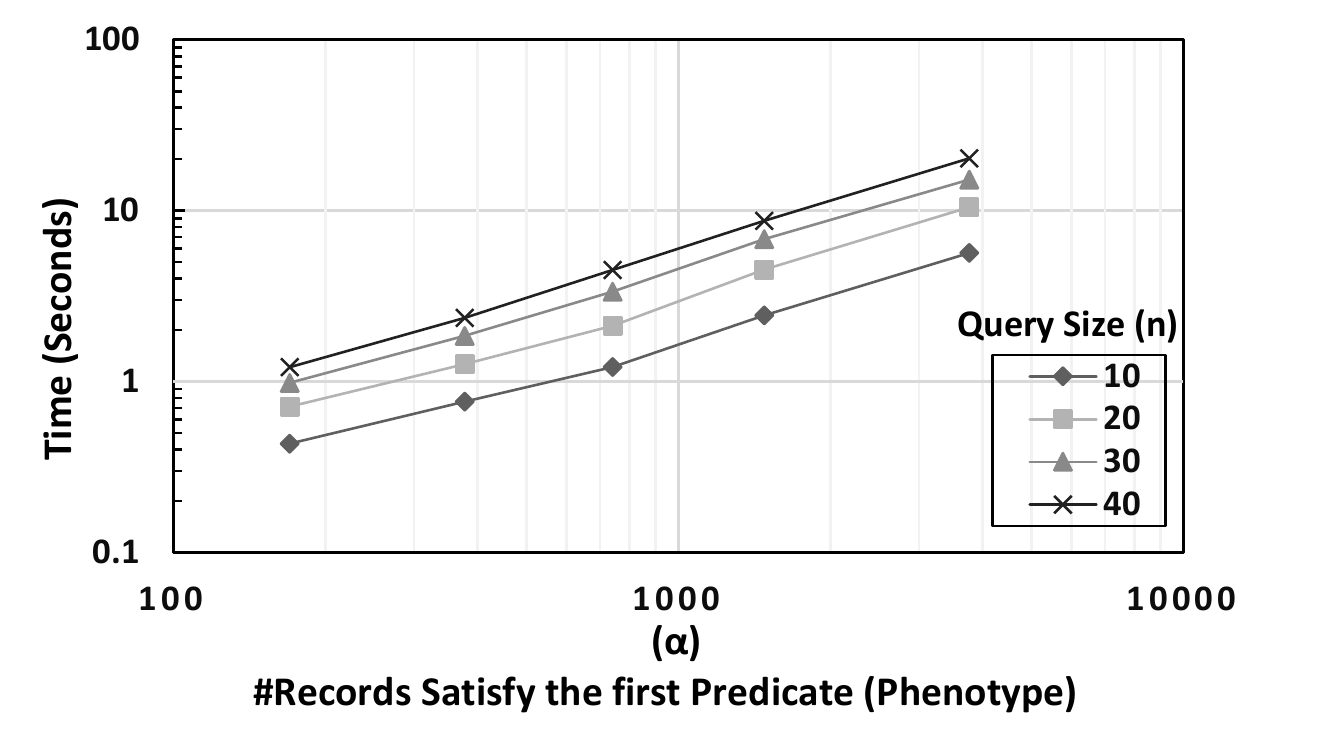}
}
\caption{Query execution time on dataset with $40,000$ records in total (different number of records satisfy the phenotype in the query).}
\label{fig:SearchCase}
\vspace{0.2cm}
\end{figure*}

The search complexity depends on the number of matched IDs to the first predicate ($\alpha$) in the query, which we selected as phenotype. Therefore, even in a large database, the search complexity is in the order of the number of returned IDs for the phenotype. Fig. \ref{fig:SearchCase} illustrates the time taken for a query to be performed on a database when the number of records that match the phenotype differ. 
This experiment shows the token generation and search time depend only on $\alpha$.

\begin{figure*}[htb]
\centering
\subfigure[\#Records with specific phenotype in different datasets]{
\includegraphics[scale=0.54]{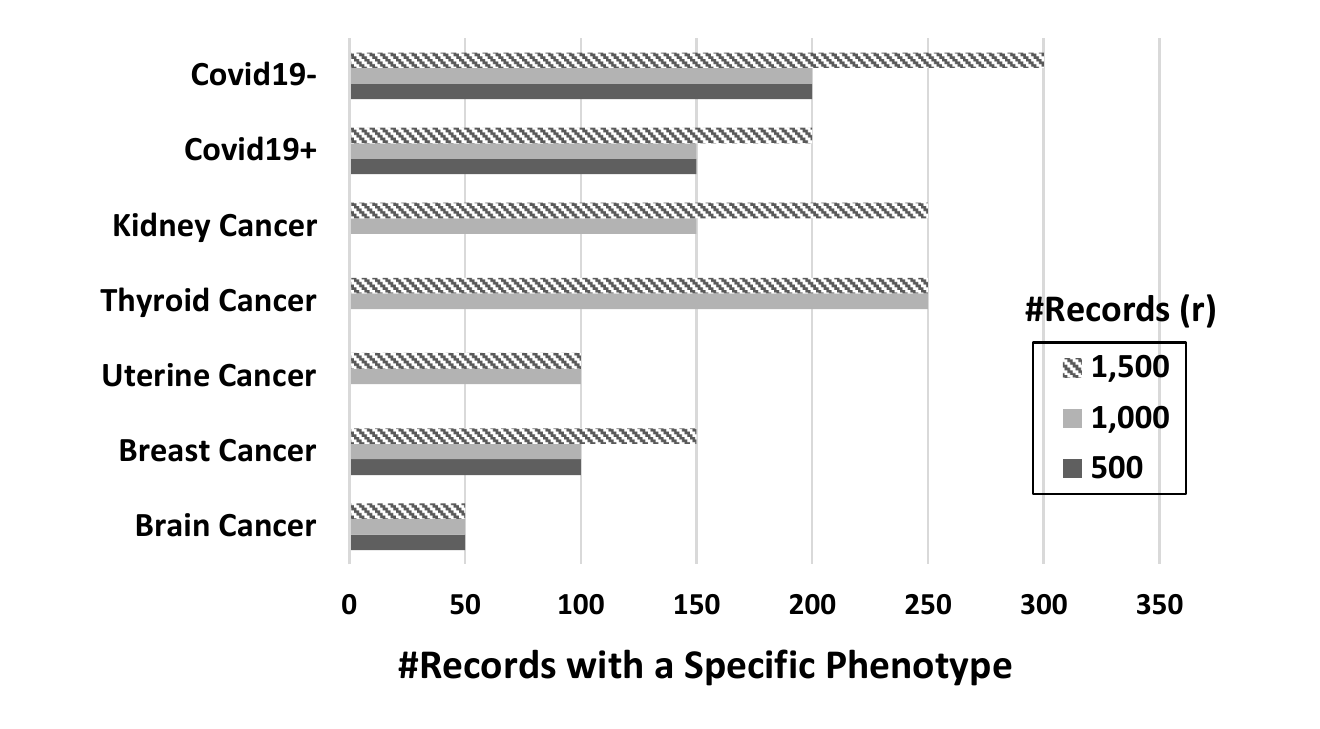}
}
\subfigure[Token Generation and Search (Data Server)]{
\includegraphics[scale=0.54]{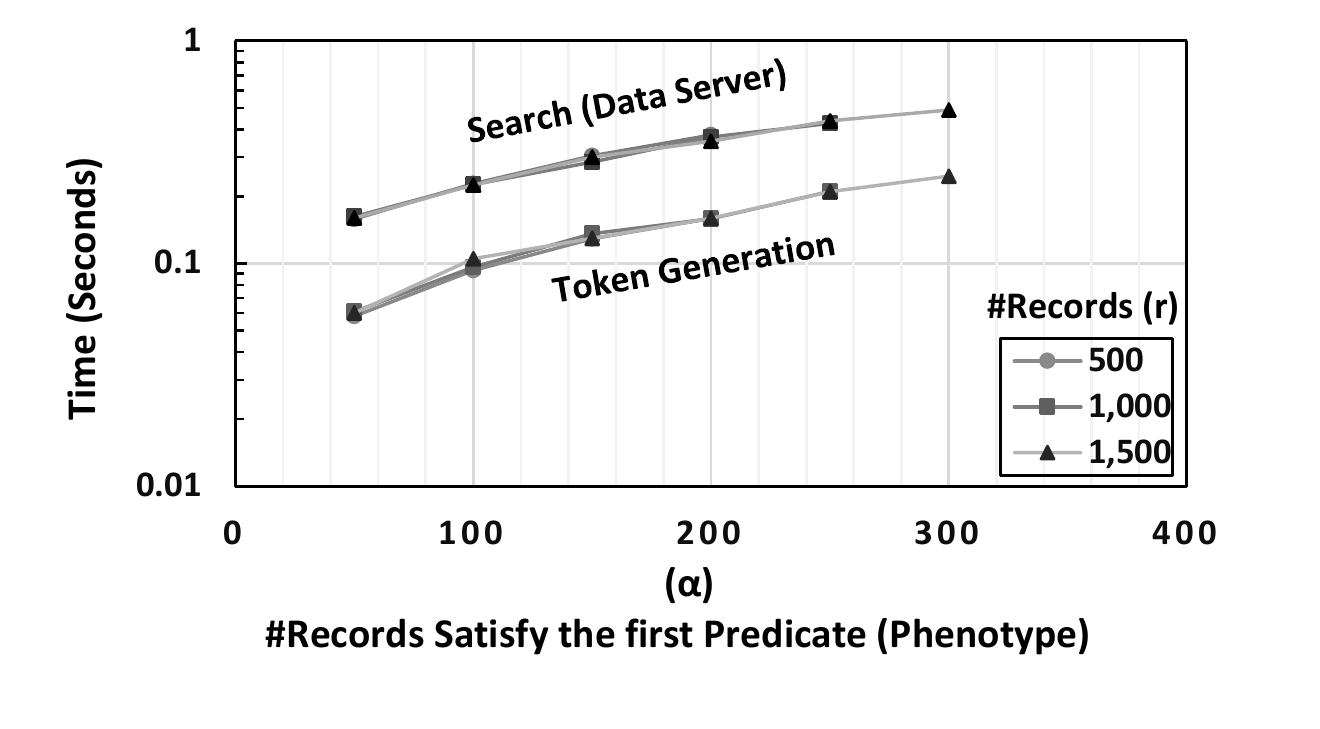}
}
\caption{Query execution time on different datasets with same number of records satisfy the phenotype in the query}
\label{fig:SearchPheno}
\vspace{0.2cm}
\end{figure*}

\begin{figure*}[htb]
\centering
\subfigure[Token Generation]{
\includegraphics[scale=0.52]{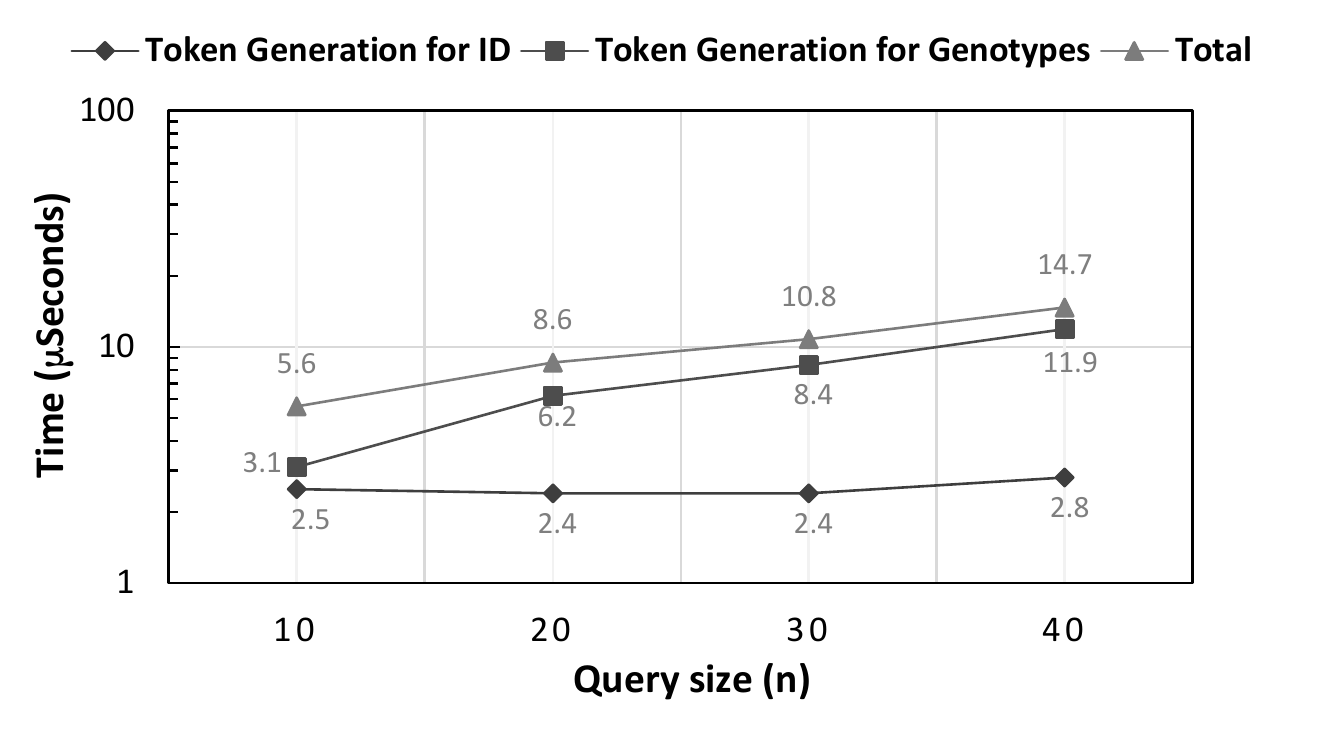}
}
\subfigure[Search (Data Server)]{
\includegraphics[scale=0.52]{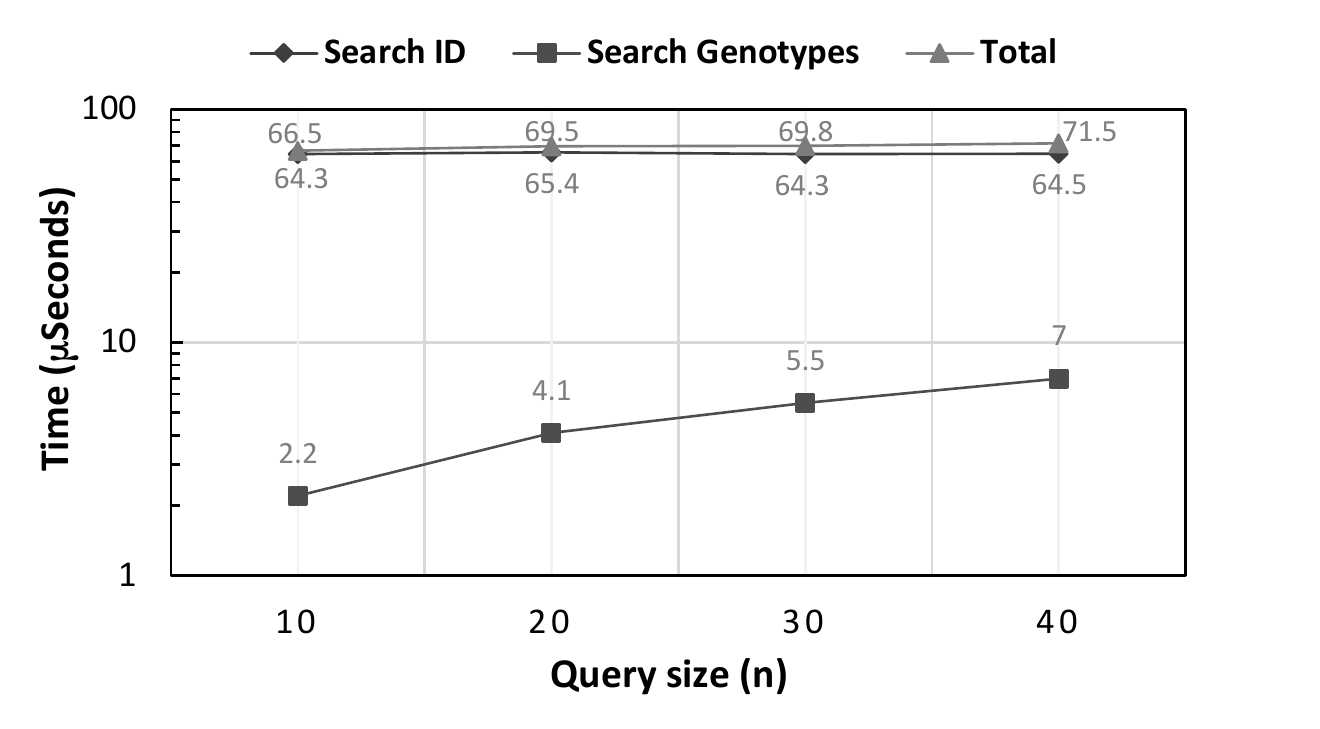}
}
\caption{$k'$-out-of-$k$ match query execution time for one patient on dataset with $5,000$ records and $1,000$ SNPs.} 
\label{fig:SearchK'}
\end{figure*}

To explicitly show the scalability of PrivGenDB, we explored the query execution time on datasets with $500$, $1,000$, and $1,500$  records, while they have the same number of IDs with particular cancer or Covid19+/-.
Fig. \ref{fig:SearchPheno}.a represents different datasets with the same number of records for some phenotypes.
Fig. \ref{fig:SearchPheno}.b shows that, for example, when the query's first predicate is thyroid cancer in $r=1,000$ and $r=1,500$ database, the search complexity is the same since both have $250$ records with this disease. Moreover, the number of records with brain cancer is $50$ in all three datasets, and it is obvious from Fig. \ref{fig:SearchPheno}.b that the token generation time and search time is the same in these three datasets.

\subsubsection{Scalability of PrivGenDB in terms of $k'$-out-of-$k$ match query for one patient}
Our experiments show that the query execution time for conducting $k'$-out-of-$k$ match query for a particular patient, when a clinician is interested, is mostly dependent on the time taken for the $\Server$ to search for the ID in $\GSet$, which is the first predicate in this query. Token generation on $\Vetter$ side and checking the genotypes on $\Server$ side depends on the query size, but the whole process takes around $70\mu$seconds (see Fig. \ref{fig:SearchK'}). 

\subsubsection{Performance comparison}
Table \ref{table:numericalcomparison} compares the query execution time of PrivGenDB with \cite{kantarcioglu2008cryptographic,canim2011secure,ghasemi2016private,hasan2018secure}.
The times of \cite{kantarcioglu2008cryptographic,canim2011secure,ghasemi2016private,hasan2018secure} are reported directly from their original paper.\footnote{We acknowledge that running the experiments on the
same machine as ours should improve the performance of \cite{kantarcioglu2008cryptographic,canim2011secure}.}
The query execution time of our model depends mostly on the number of records matching the first predicate ($\alpha$), which is the phenotype in our model, whereas it is linear to the number of records ($r$) for \cite{kantarcioglu2008cryptographic,canim2011secure,ghasemi2016private} and linear to the number of SNPs in \cite{hasan2018secure}. In comparison to \cite{hasan2018secure}, which is the only other model that supports storing the phenotype in the database, PrivGenDB takes only $5.2$ seconds to execute a query of size $50$ on our $5,000$ database. It is worth mentioning that it may take less time if the number of IDs that match a phenotype is less than that in the experiment. Whereas, \cite{hasan2018secure} takes around $130$ seconds to execute the query on the encrypted tree structure. The query execution time in \cite{nassar2017securing} is in the order of the number of records stored in the data server.

\definecolor{magentaa}{cmyk}{0.2,0.4,0,0}
\begin{table}[htb]
\caption{Comparison of query execution time on $5,000$ 
records with query sizes between $10$ and $50$ SNPs}\label{table:numericalcomparison}
\centering
\resizebox{\columnwidth}{!}{
\begin{tabular}{ccccccc}
\toprule
\multirow{2}{*}{Scheme}    & \multirow{2}{*}{Phenotype} & \multicolumn{5}{c}{Query Size}                 \\ \cmidrule{3-7} 
                           &                            & $10$      & $20$      & $30$      & $40$      & $50$      \\ \hline
\cite{kantarcioglu2008cryptographic}                        & no                         & $1260$ s  & $1270$ s  & $1285$ s  & $1290$ s  & N/A     \\
[1pt] 
\hline
\cite{canim2011secure}                      & p/n                        & $20$ s    & $40$ s    & $60$ s    & $80$ s    & $100$ s   \\[1pt] \hline
\cite{ghasemi2016private}                    & p/n                        & $29$ s    & $31$ s    & $24$ s    & $37$ s    & N/A     \\[1pt] \hline
\cite{hasan2018secure}                      & yes                        & $130.8$ s & N/A     & N/A     & N/A     & $128.7$ s \\[1pt] \hline
\cite{new2020}                      & yes                        & $163.92$ s & N/A     & N/A     & N/A     & $143.9$ s \\[1pt] \hline
\multirow{2}{*}{PrivGenDB} & \multirow{2}{*}{yes}       &$\#$  $1.3$ s  & $2.3$ s  & $3.5$ s  & $4.3$ s  & $5.2$ s   \\[1pt] \cline{3-7} 
                           &                            & $\S$  $72.3$ $\mu$s & $78.2$ $\mu$s & $80.7$ $\mu$s & $86.4$ $\mu$s & $88.5$ $\mu$s \\[1pt] \bottomrule
                           
 \multicolumn{7}{p{10.5cm}}{p/n: positive/negative signs to show if a record has or does not have a particular disease; $\#$: This row is related to the count and Boolean queries execution time; $\S$: This row is related to the $k'$-out-of-$k$ match query execution time for a particular random patient on database.}\\                          
\end{tabular}}
\end{table}

\subsection{Communication Overhead}
\begin{figure}[htb]
\centering

\includegraphics[scale=0.52]{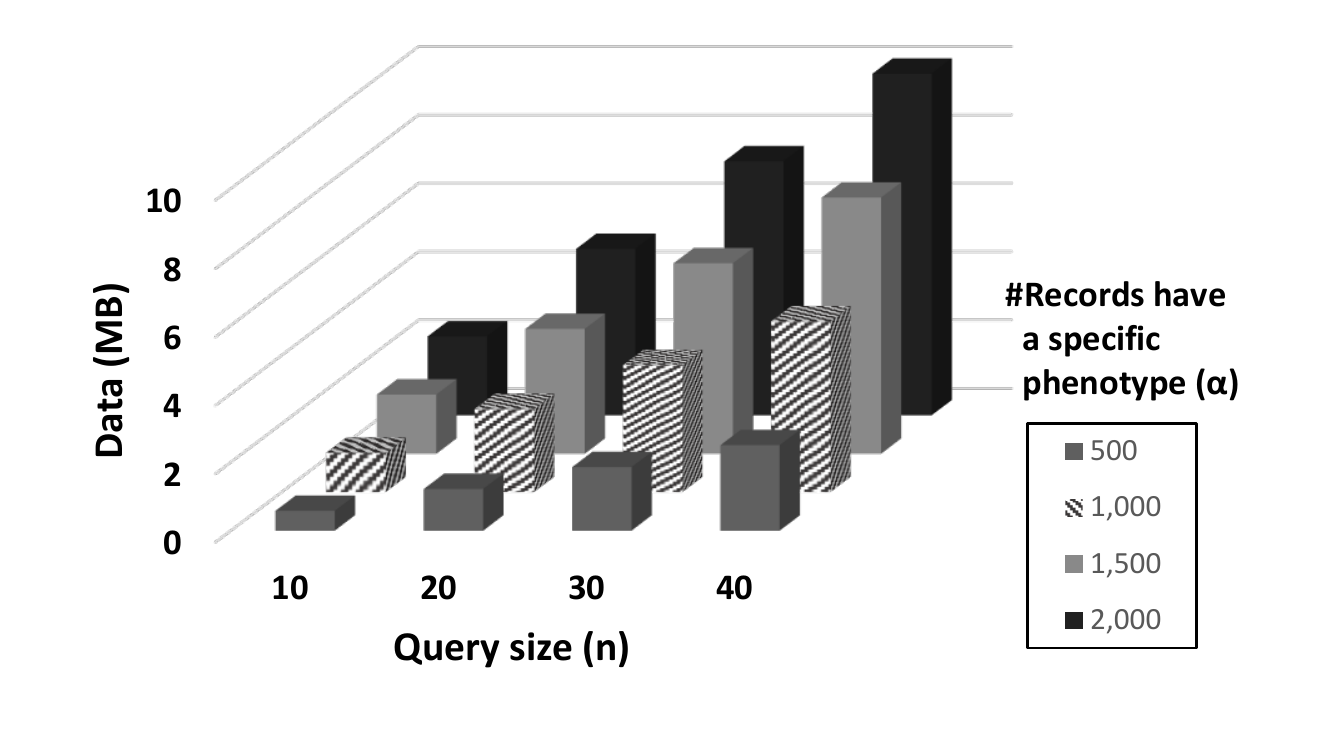}

\caption{Total communication cost for different query executions on dataset with different $\alpha$.}
\label{fig:communicationoverhead}
\end{figure}

The amount of data transferred between the $\Server$ and the $\Vetter$ is depicted in Fig. \ref{fig:communicationoverhead}.
This is the data amount to execute the query with different query sizes while $\alpha$ changes. 
This overhead increases if the query size ($n$) or the number of records match the phenotype ($\alpha$) increase, which defines the tokens generated by $\Vetter$ and transmitted to the $\Server$ for searching. Both the query execution time and communication cost generally depend on these two parameters, linearly.
Query privacy is not provided in \cite{nassar2017securing}, and \cite{ghasemi2016private} does not encrypt the query.
This overhead increases linearly if the number of nodes need to be accessed to perform a query increase in \cite{hasan2018secure}, and increases linearly by the number of records in \cite{kantarcioglu2008cryptographic,canim2011secure}.

\section{Conclusion and Future works}
In this paper, we proposed a new secure and efficient model, named PrivGenDB, for outsourcing Single Nucleotide Polymorphism (SNP)-Phenotype data to the cloud server. 
To the best of our knowledge, PrivGenDB is the first model that ensures the confidentiality of shared SNP-phenotype data by employing searchable symmetric encryption (SSE) and makes the computation/query process efficient.
PrivGenDB constructs an inverted index using a novel encoding mechanism for genotypes, phenotypes, gender, and ethnicity; then encrypts it. By employing SSE, the vetter can generate tokens, and the data server can execute different query operations.
We have compared PrivGenDB to the state-of-the-art privacy-preserving query executions on SNP-Phenotype data both analytically and numerically. Our experimental results on real and synthetic data demonstrate that the proposed model outperforms the existing schemes in terms of query execution time.

Similar to the other existing related works, PrivGenDB has some leakages that might lead to vulnerabilities to some attacks (e.g., statistical attacks). This can be avoided by countermeasures like adding noise to hide the search frequencies. We leave studying them as a future work.
Finally, PrivGenDB is flexible to support consent, which we can further take into consideration as another promising research direction.




\begin{thebibliography}{10}
\providecommand{\url}[1]{#1}
\csname url@samestyle\endcsname
\providecommand{\newblock}{\relax}
\providecommand{\bibinfo}[2]{#2}
\providecommand{\BIBentrySTDinterwordspacing}{\spaceskip=0pt\relax}
\providecommand{\BIBentryALTinterwordstretchfactor}{4}
\providecommand{\BIBentryALTinterwordspacing}{\spaceskip=\fontdimen2\font plus
\BIBentryALTinterwordstretchfactor\fontdimen3\font minus
  \fontdimen4\font\relax}
\providecommand{\BIBforeignlanguage}[2]{{%
\expandafter\ifx\csname l@#1\endcsname\relax
\typeout{** WARNING: IEEEtran.bst: No hyphenation pattern has been}%
\typeout{** loaded for the language `#1'. Using the pattern for}%
\typeout{** the default language instead.}%
\else
\language=\csname l@#1\endcsname
\fi
#2}}
\providecommand{\BIBdecl}{\relax}
\BIBdecl

\bibitem{hasan2018secure}
M.~Z. Hasan, M.~R. Mahdi, M.~Sadat, and N.~Mohammed, ``Secure count query on
  encrypted genomic data,'' \emph{Journal of biomedical informatics}, vol.~81,
  pp. 41--52, 2018.

\bibitem{ghasemi2016private}
R.~Ghasemi, M.~A. Aziz, N.~Mohammed, M.~H. Dehkordi, and X.~Jiang, ``Private
  and efficient query processing on outsourced genomic databases,'' \emph{IEEE
  journal of biomedical and health informatics}, vol.~21, no.~5, pp.
  1466--1472, 2016.

\bibitem{nassar2017securing}
M.~Nassar, Q.~Malluhi, M.~Atallah, and A.~Shikfa, \emph{Securing aggregate
  queries for dna databases}.\hskip 1em plus 0.5em minus 0.4em\relax IEEE
  Trans. on Cloud Computing, 2017.

\bibitem{kantarcioglu2008cryptographic}
M.~Kantarcioglu, W.~Jiang, Y.~Liu, and B.~Malin, ``A cryptographic approach to
  securely share and query genomic sequences,'' \emph{IEEE Trans Inform Technol
  Biomed}, vol.~12, no.~5, pp. 606--617, September 2008.

\bibitem{canim2011secure}
M.~Canim, M.~Kantarcioglu, and B.~Malin, ``Secure management of biomedical data
  with cryptographic hardware,'' \emph{IEEE Trans Inform Technol Biomed},
  vol.~16, no.~1, pp. 166--175, January 2012.

\bibitem{chenghong2017scotch}
W.~Chenghong, Y.~Jiang, N.~Mohammed, F.~Chen, X.~Jiang, M.~A. Aziz, M.~Sadat,
  and S.~Wang, \emph{Scotch: Secure counting of encrypted genomic data using a
  hybrid approach}.\hskip 1em plus 0.5em minus 0.4em\relax 1744: In {\em AMIA
  Annual Symposium Proceedings}, 2017, vol. page.

\bibitem{NationalHumanGenomeResearchInstitute}
\BIBentryALTinterwordspacing
N.~H. G.~R. Institute, \emph{Phenotype}.\hskip 1em plus 0.5em minus 0.4em\relax
  \url. [Online]. Available:
  \url{https://www.genome.gov/genetics-glossary/Phenotype}
\BIBentrySTDinterwordspacing

\bibitem{NationalHumanGenomeResearchInstitute1}
G.~Gibson, ``Population genetics and gwas: a primer,'' \emph{PLoS Biol},
  vol.~16, 2018.

\bibitem{cash2013highly}
D.~Cash, S.~Jarecki, C.~Jutla, H.~Krawczyk, M.-C. Ro{\c{s}}u, and M.~Steiner,
  ``Highly-scalable searchable symmetric encryption with support for boolean
  queries,'' in \emph{{\em Annual cryptology conference}}, 2013, pp. 353--373.

\bibitem{UKbiobank1}
C.~Sudlow, J.~Gallacher, N.~Allen, V.~Beral, P.~Burton, J.~Danesh, P.~Downey,
  P.~Elliott, J.~Green, M.~Landray \emph{et~al.}, ``Uk biobank: an open access
  resource for identifying the causes of a wide range of complex diseases of
  middle and old age,'' \emph{Plos med}, vol.~12, no.~3, p. e1001779, 2015.

\bibitem{UKbiobank2}
C.~Bycroft, C.~Freeman, D.~Petkova, G.~Band, L.~T. Elliott, K.~Sharp,
  A.~Motyer, D.~Vukcevic, O.~Delaneau, J.~O’Connell \emph{et~al.}, ``The uk
  biobank resource with deep phenotyping and genomic data,'' \emph{Nature},
  vol. 562, no. 7726, pp. 203--209, 2018.

\bibitem{NIH}
``Nih all of us research program.'' \url{https://allofus.nih.gov/}, no date.

\bibitem{MVP}
``Million veteran program (mvp).'' \url{https://www.mvp.va.gov/}, no date.

\bibitem{personalized1}
P.~M. Visscher, N.~R. Wray, Q.~Zhang, P.~Sklar, M.~I. McCarthy, M.~A. Brown,
  and J.~Yang, ``10 years of gwas discovery: biology, function, and
  translation,'' \emph{The American Journal of Human Genetics}, vol. 101,
  no.~1, pp. 5--22, 2017.

\bibitem{personalized2}
G.~S. Ginsburg and K.~A. Phillips, ``Precision medicine: from science to
  value,'' \emph{Health Affairs}, vol.~37, no.~5, pp. 694--701, 2018.

\bibitem{cloudsecurity1}
Y.~Al-Issa, M.~A. Ottom, and A.~Tamrawi, ``ehealth cloud security challenges: a
  survey,'' \emph{Journal of healthcare engineering}, vol. 2019, 2019.

\bibitem{cloudsecurity2}
T.~Ermakova, B.~Fabian, and R.~Zarnekow, ``Improving individual acceptance of
  health clouds through confidentiality assurance,'' \emph{Applied clinical
  informatics}, vol.~7, no.~4, p. 983, 2016.

\bibitem{cloudsecurity3}
M.-H. Kuo, ``Opportunities and challenges of cloud computing to improve health
  care services,'' \emph{Journal of medical Internet research}, vol.~13, no.~3,
  p. e67, 2011.

\bibitem{privacy1}
Y.~Erlich and A.~Narayanan, ``Routes for breaching and protecting genetic
  privacy,'' \emph{Nature Reviews Genetics}, vol.~15, no.~6, pp. 409--421,
  2014.

\bibitem{privacy2}
Y.~Erlich, J.~B. Williams, D.~Glazer, K.~Yocum, N.~Farahany, M.~Olson,
  A.~Narayanan, L.~D. Stein, J.~A. Witkowski, and R.~C. Kain, ``Redefining
  genomic privacy: trust and empowerment,'' \emph{PLoS Biol}, vol.~12, no.~11,
  p. e1001983, 2014.

\bibitem{naveed2015privacy}
M.~Naveed, E.~Ayday, E.~W. Clayton, J.~Fellay, C.~A. Gunter, J.~P. Hubaux,
  B.~A. Malin, and X.~Wang, ``Privacy in the genomic era,'' \emph{ACM Computing
  Surveys (CSUR)}, vol.~48, no.~1, pp. 1--44, 2015.

\bibitem{HIPAA}
``Guide to hipaa privacy rule and compliance,''
  \url{http://www.hipaa-101.com/}, no date.

\bibitem{BigDataforHealth}
J.~Andreu-Perez, C.~C.~Y. Poon, R.~D. Merrifield, S.~T.~C. Wong, and G.~Z.
  Yang, ``Big data for health,'' \emph{IEEE Journal of Biomedical and Health
  Informatics}, vol.~19, no.~4, pp. 1193--1208, July 2015.

\bibitem{berger2019emerging}
B.~Berger and H.~Cho, ``Emerging technologies towards enhancing privacy in
  genomic data sharing,'' \emph{Genome Biology}, vol.~20, p.~1, December 2019.

\bibitem{gymrek2013identifying}
M.~Gymrek, A.~L. McGuire, D.~Golan, E.~Halperin, and Y.~Erlich, ``Identifying
  personal genomes by surname inference,'' \emph{Science}, vol. 339, no. 6117,
  pp. 321--324, January 2013.

\bibitem{new2018}
L.~Chen, M.~M. Aziz, N.~Mohammed, , and X.~Jiang, ``Secure large-scale genome
  data storage and query,'' \emph{Computer methods and programs in
  biomedicine}, vol. 165, pp. 129--137, 2018.

\bibitem{new2020}
M.~S.~R. Mahdi, M.~N. Sadat, N.~Mohammed, and X.~Jiang, ``Secure count query on
  encrypted heterogeneous data,'' in \emph{2020 IEEE Intl Conf on
  (DASC/PiCom/CBDCom/CyberSciTech)}.\hskip 1em plus 0.5em minus 0.4em\relax
  IEEE, 2020, pp. 548--555.

\bibitem{SEsurvey}
R.~Zhang, R.~Xue, , and L.~Liu, ``Searchable encryption for healthcare clouds:
  a survey,'' \emph{IEEE Transactions on Services Computing}, vol.~11, no.~6,
  pp. 978--996, 2017.

\bibitem{SSE20181}
S.~Lai, S.~Patranabis, A.~Sakzad, J.~K. Liu, D.~Mukhopadhyay, R.~Steinfeld,
  S.~F. Sun, D.~Liu, and C.~Zuo, ``Result pattern hiding searchable encryption
  for conjunctive queries,'' \emph{Proceedings of the 2018 ACM {CCS}}, pp.
  745--762, 2018.

\bibitem{SSE20182}
S.~F. Sun, X.~Yuan, J.~K. Liu, R.~Steinfeld, A.~Sakzad, V.~Vo, and S.~Nepal,
  ``Practical backward-secure searchable encryption from symmetric puncturable
  encryption,'' \emph{Proceedings of the 2018 ACM CCS}, vol. 763, 2018.

\bibitem{SSE2016}
S.~K. Kermanshahi, J.~K. Liu, R.~Steinfeld, S.~Nepal, S.~Lai, R.~Loh, and
  C.~Zuo, \emph{Multi-client Cloud-based Symmetric Searchable
  Encryption}.\hskip 1em plus 0.5em minus 0.4em\relax IEEE Trans. Dependable
  Secure Comput, 2019.

\bibitem{SSE2020}
S.~K. Kermanshahi, S.~F. Sun, J.~K. Liu, R.~Steinfeld, S.~Nepal, W.~F. Lau, and
  M.~H. Au., \emph{Geometric range search on encrypted data with
  forward/backward security}.\hskip 1em plus 0.5em minus 0.4em\relax IEEE
  Trans. Dependable Secure Comput, 2020.

\bibitem{SSE2019new}
S.~Lai, X.~Yuan, S.~Sun, J.~K. Liu, Y.~Liu, and D.~Liu, ``Graphse$^2$: An
  encrypted graph database for privacy-preserving social search,'' in
  \emph{Proceedings of the 2019 ACM Asia Conference on Computer and
  Communications Security}, 2019, pp. 41--54.

\bibitem{SSE2021new}
S.~Sun, R.~Steinfeld, S.~Lai, X.~Yuan, A.~Sakzad, J.~K. Liu, S.~Nepal, and
  D.~Gu, ``Practical non-interactive searchable encryption with forward and
  backward privacy,'' \emph{NDSS. The Internet Society}, 2021.

\bibitem{substringsearch}
Y.~Uchide and N.~Kunihiro, ``Searchable symmetric encryption capable of
  searching for an arbitrary string,'' \emph{Security and Communication
  Networks}, vol.~9, no.~12, pp. 1726--1736, 2016.

\bibitem{rangequery}
W.~Sun, N.~Zhang, W.~Lou, and Y.~T. Hou, ``When gene meets cloud: Enabling
  scalable and efficient range query on encrypted genomic data,'' in \emph{IEEE
  INFOCOM 2017-IEEE Conference on Computer Communications}.\hskip 1em plus
  0.5em minus 0.4em\relax IEEE, 2017, pp. 1--9.

\bibitem{chen2017princess}
F.~Chen, S.~Wang, X.~Jiang, S.~Ding, Y.~Lu, J.~Kim, S.~C. Sahinalp, C.~Shimizu
  \emph{et~al.}, ``Princess: Privacy-protecting rare disease international
  network collaboration via encryption through software guard extensions,''
  \emph{Bioinformatics}, vol.~33, no.~6, pp. 871--878, 2017.

\bibitem{prf}
J.~Katz and Y.~Lindell, \emph{Introduction to modern cryptography book}.\hskip
  1em plus 0.5em minus 0.4em\relax In {\em CRC press}, 2020.

\bibitem{bf}
B.~H. Bloom, ``Space/time trade-offs in hash coding with allowable errors,''
  \emph{Communications of the ACM}, vol.~13, pp. 422--426, 1970.

\bibitem{Redis}
\BIBentryALTinterwordspacing
R.~Labs, ``Redis,'' vol. 2017. [Online]. Available: \url{https://redis.io}
\BIBentrySTDinterwordspacing

\bibitem{BF1}
\BIBentryALTinterwordspacing
A.~Nikitin, \emph{Bloom Filter Scala}, vol. 2017. [Online]. Available:
  \url{https://alexandrnikitin.github.io/blog/bloom-filter-for-scala/}
\BIBentrySTDinterwordspacing

\bibitem{JPBC}
A.~D. Caro and V.~Iovino, \emph{jpbc: Java pairing based cryptography}.\hskip
  1em plus 0.5em minus 0.4em\relax pages 850-855. IEEE: In {\em ISCC 2011},
  2011.

\bibitem{PGP}
\hskip 1em plus 0.5em minus 0.4em\relax The Personal Genome Project: Harvard
  Medical School. \url{}, title = {PersonalGenomes.org}, url =
  {https://pgp.med.harvard.edu/data}.

\end{thebibliography}


\begin{thebibliography}{99}

\bibitem{UKbiobank1}
C.~Sudlow, J.~Gallacher, N.~AllenSudlow, et al.
\newblock UK biobank: an open access resource for identifying the causes of a wide range of complex diseases of middle and old age.
\newblock {\em PLoS Med.}, 12, e1001779, Mar. 2015.

\bibitem{UKbiobank2}
C.~Bycroft, C.~Freeman, D.~Petkova, et al.
\newblock The UK Biobank resource with deep phenotyping and genomic data.
\newblock {\em Nature}, 2018.

\bibitem{NIH}
NIH All of Us Research Program.
\newblock \url{https://allofus.nih.gov/}.

\bibitem{MVP}
Million Veteran Program (MVP).
\newblock \url{https://www.mvp.va.gov/}.




\bibitem{personalized1}
 PM.~Visscher, NR.~Wray, Q.~Zhang, P.~Sklar, MI.~McCarthy, MA.~Brown, J.~Yang.
\newblock 10 Years of GWAS Discovery:
Biology, Function, and Translation.
\newblock {\em The American Journal of Human Genetics}, 101:5–22, 2017.

\bibitem{personalized2}
G.~Ginsburg, KA.~Phillips.
\newblock Precision medicine: from science to value.
\newblock {\em Health Aff (Millwood)}, pages 694-701, 2018.




\bibitem{cloudsecurity1}
Y.~Al-Issa, M.A.~Ottom, A.~Tamrawi.
\newblock ehealth cloud security challenges: A survey.
\newblock {\em Journal of healthcare engineering}, 2019.

\bibitem{cloudsecurity2}
T.~Ermakova, B.~Fabian, R.~Zarnekow.
\newblock Improving individual acceptance of health clouds through confidentiality assurance.
\newblock {\em Appl Clin Inform.}, 2016;7:983–93.
 
\bibitem{cloudsecurity3}
A.M.-H.~Kuo.
\newblock Opportunities and challenges of cloud computing to improve health
  care services.
\newblock {\em Journal of Medical Internet Research}, 13(3):e67, Sep. 2011.

\bibitem{privacy1}
Y.~Erlich and A.~Narayanan.
\newblock Routes for breaching and protecting genetic privacy.
\newblock {\em Nature reviews. Genetics}, 2014. 15(6): 409–21.

\bibitem{privacy2}
Y.~Erlich, JB.~Williams, D.~Glazer, et al.
\newblock Redefining genomic privacy: trust and empowerment.
\newblock {\em PLoS Biol.}, 2014; 12: e1001983.


\bibitem{naveed2015privacy}
M.~Naveed, E.~Ayday, E.W.~Clayton, J.~Fellay, C.A.~Gunter,
  J.P.~Hubaux, B.A.~Malin, and X.~Wang.
\newblock Privacy in the genomic era.
\newblock {\em ACM Computing Surveys (CSUR)}, 48(1):1--44, 2015.
  
\bibitem{HIPAA}
Guide to hipaa privacy rule and compliance,
\newblock  HIPAA.
\newblock \url{http://www.hipaa-101.com/}.  
  
  
\bibitem{BigDataforHealth}
J.~Andreu-Perez, C.C.Y.~Poon, R.D.~Merrifield, S.T.C.~Wong,
and G.Z.~Yang.
\newblock Big data for health.
\newblock {\em IEEE Journal of Biomedical and Health Informatics},
  19(4):1193--1208, Jul. 2015.

\bibitem{berger2019emerging}
B.~Berger and H.~Cho.
\newblock Emerging technologies towards enhancing privacy in genomic data
  sharing.
\newblock {\em Genome Biology}, 20(1), Dec 2019.


\bibitem{gymrek2013identifying}
M.~Gymrek, A.~L. McGuire, D.~Golan, E.~Halperin, and Y.~Erlich.
\newblock Identifying personal genomes by surname inference.
\newblock {\em Science}, 339(6117):321--324, Jan. 2013.

\bibitem{hasan2018secure}
M.Z.~Hasan, Md.S.R.~Mahdi, Md.N.~Sadat, and N.~Mohammed.
\newblock Secure count query on encrypted genomic data.
\newblock {\em Journal of biomedical informatics}, 81:41--52, 2018.

\bibitem{kantarcioglu2008cryptographic}
M.~Kantarcioglu, W.~Jiang, Y.~Liu, and B.~Malin.
\newblock A cryptographic approach to securely share and query genomic
  sequences.
\newblock {\em IEEE Trans Inform Technol Biomed},
  12(5):606--617, Sep. 2008.

\bibitem{canim2011secure}
M.~Canim, M.~Kantarcioglu, and B.~Malin.
\newblock Secure management of biomedical data with cryptographic hardware.
\newblock {\em IEEE Trans Inform Technol Biomed},
  16(1):166--175, Jan. 2012.

\bibitem{ghasemi2016private}
R.~Ghasemi, Md.M.~Al~Aziz, N.~Mohammed, M.H.~Dehkordi, and
  X.~Jiang.
\newblock Private and efficient query processing on outsourced genomic
  databases.
\newblock {\em IEEE journal of biomedical and health informatics},
  21(5):1466--1472, 2016.
  
\bibitem{nassar2017securing}
M.~Nassar, Q.~Malluhi, M.~Atallah, and A.~Shikfa.
\newblock Securing aggregate queries for dna databases.
\newblock {\em IEEE Trans. on Cloud Computing}, 2017.

\bibitem{chenghong2017scotch}
W.~Chenghong, Y.~Jiang, N.~Mohammed, F.~Chen, X.~Jiang,
  Md.M.~Al~Aziz, Md.N.~Sadat, and S.~Wang.
\newblock Scotch: Secure counting of encrypted genomic data using a hybrid
  approach.
\newblock In {\em AMIA Annual Symposium Proceedings}, volume 2017, page 1744.


\bibitem{new2018}
L.~Chen, Md M.~Aziz, N.~Mohammed, X.~Jiang.
\newblock Secure large-scale genome data storage and query.
\newblock {\em Computer methods and programs in biomedicine}, Elsevier, Vol. 165, pages 129-137, 2018.

\bibitem{new2020}
Md.S.R.~Mahdi, Md.N.~Sadat, and N.~Mohammed, X.~Jiang.
\newblock Secure Count Query on Encrypted Heterogeneous Data.
\newblock {\em 2020 Intl Conf on \remove{Dependable, Autonomic and Secure Computing, Intl Conf on Pervasive Intelligence and Computing, Intl Conf on Cloud and Big Data Computing, Intl Conf on Cyber Science and Technology Congress} (DASC/PiCom/CBDCom/CyberSciTech)}, pages 548-555, 2020.



\bibitem{SEsurvey}
R.~Zhang, R.~Xue, L.~Liu.
\newblock Searchable encryption for healthcare clouds: a survey.
\newblock {\em IEEE Transactions on Services Computing}, Vol. 11, No. 6, pages 978-996, 2017.


\bibitem{SSE20181}
S.~Lai, S.~Patranabis, A.~Sakzad, J.K.~Liu, D.~Mukhopadhyay, R.~Steinfeld, S.~F.~Sun, D.~Liu, C.~Zuo.
\newblock Result pattern hiding searchable encryption for conjunctive queries.
\newblock {\em Proceedings of the 2018 ACM {CCS}}, pages 745-762, 2018.

\bibitem{SSE20182}
S.~F.~Sun, X.~Yuan, J.K.~Liu, R.~Steinfeld, A.~Sakzad, V.~Vo, S.~Nepal.
\newblock Practical backward-secure searchable encryption from symmetric puncturable encryption.
\newblock {\em Proceedings of the 2018 ACM CCS}, pages 763–780 , 2018.



\bibitem{SSE2016}
S.~Kasra Kermanshahi, J.K.~Liu, R.~Steinfeld, S.~Nepal, S.~Lai, R.~Loh, C.~Zuo.
\newblock Multi-client Cloud-based Symmetric Searchable Encryption.
\newblock {\em IEEE Trans. Dependable Secure Comput.}, 2019.

\bibitem{SSE2020}
S.~Kasra~Kermanshahi, S.F.~Sun, J.K.~Liu, R.~Steinfeld, S.~Nepal, W.F.~Lau, M.H.~Au.
\newblock Geometric range search on encrypted data with forward/backward security.
\newblock {\em IEEE TDSC.}, 2020.



\bibitem{SSE2019new}
Sh.~Lai, X.~Yuan, Sh.~Sun, J.K.~Liu, Y.~Liu, D.~Liu.
\newblock GraphSE$^2$: An Encrypted Graph Database for Privacy-Preserving Social Search.
\newblock {\em Proceedings of the ACM Asia CCS.}, 2019.


\bibitem{SSE2021new}
Sh.~Sun, R.~Steinfeld, Sh.~Lai, X.~Yuan, A.~Sakzad, J.K.~Liu, S.~Nepal, D.~Gu.
\newblock Practical Non-Interactive Searchable Encryption with Forward and Backward Privacy.
\newblock {\em NDSS Symposium.}, 2021.



\bibitem{substringsearch}
Y.~Uchide, N.~Kunihiro.
\newblock Searchable symmetric encryption capable of searching for an arbitrary string.
\newblock {\em Security and Communication Networks.}, Vol. 9, No. 12, pages 1726-1736, 2016.

\bibitem{rangequery}
W.~Sun, N.~Zhang, W.~Lou, Y Th.~Hou.
\newblock When gene meets cloud: Enabling scalable and efficient range query on encrypted genomic data.
\newblock {\em IEEE INFOCOM.}, pages 1-9, 2017.

\bibitem{cash2013highly}
D.~Cash, S.~Jarecki, C.~Jutla, H.~Krawczyk,
  M-C.~Ro{\c{s}}u, and M.~Steiner.
\newblock Highly-scalable searchable symmetric encryption with support for
  Boolean queries.
\newblock In {\em Annual cryptology conference}, pages 353--373. Springer,
  2013.  
















\remove{\bibitem{NationalHumanGenomeResearchInstitute}
National Human Genome~Research Institute.
\newblock Phenotype.
\newblock \url{https://www.genome.gov/genetics-glossary/Phenotype}.

\bibitem{NationalHumanGenomeResearchInstitute1}
G.~Gibson.
\newblock Population genetics and GWAS: a primer.
\newblock {\em PLoS Biol.}, 2018; 16:e2005485.


\bibitem{chen2017princess}
F.~Chen, S.~Wang, X.~Jiang, S.~Ding, Y.~Lu, J.~Kim, S.C.~Sahinalp, C.~Shimizu, et~al.
\newblock Princess: Privacy-protecting rare disease international network
  collaboration via encryption through software guard extensions.
\newblock {\em Bioinformatics}, 33(6):871--878, 2017.
}


\bibitem{prf}
J.~Katz, Y.~Lindell.
\newblock Introduction to modern cryptography book.
\newblock {\em CRC press}, 2020.

\bibitem{bf}
B H.~Bloom.
\newblock Space/time trade-offs in hash coding with allowable errors.
\newblock {\em Communications of the ACM}, Vol. 13, pages 422-426, 1970.





\bibitem{Redis}
Redis Labs.
\newblock Redis.
\newblock \url{https://redis.io}, 2017.

\bibitem{BF1}
A.~Nikitin.
\newblock Bloom Filter Scala.
\newblock \url{https://alexandrnikitin.github.io/blog/bloom-filter-for-scala/}, 2017.


\bibitem{JPBC}
Angelo De Caro and Vincenzo Iovino.
\newblock jpbc: Java pairing based
cryptography. 
\newblock  In {\em ISCC 2011}, pages 850–855. IEEE, 2011.

\bibitem{PGP}
PersonalGenomes.org.
\newblock  The Personal Genome Project: Harvard Medical School.
\newblock \url{https://pgp.med.harvard.edu/data}, (accessed Dec 2020).

\end{thebibliography}
\end{document}